\documentclass[11pt]{article}
	
	\newcommand{\blind}{0}
	
    \makeatletter
    \renewcommand\section{\@startsection {section}{1}{\z@}%
                                       {-3.5ex \@plus -1ex \@minus -.2ex}%
                                       {2.3ex \@plus.2ex}%
                                       {\normalfont\fontfamily{phv}\fontsize{14}{17}\bfseries}}
    \renewcommand\subsection{\@startsection{subsection}{2}{\z@}%
                                         {-3.25ex\@plus -1ex \@minus -.2ex}%
                                         {1.5ex \@plus .2ex}%
                                         {\normalfont\fontfamily{phv}\fontsize{12}{15}\bfseries}}
    \renewcommand\subsubsection{\@startsection{subsubsection}{3}{\z@}%
                                        {-3.25ex\@plus -1ex \@minus -.2ex}%
                                         {1.5ex \@plus .2ex}%
                                         {\normalfont\normalsize\fontfamily{phv}\fontsize{12}{14}\selectfont}}
    \makeatother
	
	\usepackage{xcolor}
	\usepackage{url} 
	\usepackage{amsthm,amsmath,amsfonts,amssymb,amsbsy}
    \usepackage[authoryear]{natbib}
    \usepackage[colorlinks,citecolor=blue,urlcolor=blue]{hyperref}
    \usepackage{algorithm2e}
    \usepackage{graphicx,psfrag,epsf}
    \usepackage{enumerate}
    \usepackage{placeins}
    \graphicspath{ {./images/} }
\usepackage{tabularx,booktabs}
\usepackage{diagbox}
    \usepackage{multirow}
    \usepackage[createShortEnv]{proof-at-the-end}
    
\newtheorem{thm}{Theorem}

\newtheorem{lemma}{Lemma}
\newtheorem{cor}{Corollary}

\theoremstyle{remark}
\newtheorem{defin}{Definition}

\SetAlgoCaptionSeparator{}
\SetAlCapNameFnt{\scshape}


\begin{document}
		
		\def\spacingset#1{\renewcommand{\baselinestretch}%
			{#1}\small\normalsize} \spacingset{1}
		
		\if0\blind
		{
    			\title{\bf \emph{Knowledge Distillation Decision Tree for Unravelling Black-box Machine Learning Models}}
			\author{Xuetao Lu $^a$ and J. Jack Lee $^{a*}$ \\
			$^a$ Department of Biostatistics, The University of Texas MD Anderson Cancer Center \\
			$^*$ Author for correspondence: jjlee@mdanderson.org}
			\date{}
			\maketitle
		} \fi
		
		\if1\blind
		{

            \title{\bf \emph{IISE Transactions} \LaTeX \ Template}
			\author{Author information is purposely removed for double-blind review}
			
\bigskip
			\bigskip
			\bigskip
			\begin{center}
				{\LARGE\bf \emph{IISE Transactions} \LaTeX \ Template}
			\end{center}
			\medskip
		} \fi
		\bigskip
		
	\begin{abstract}
Machine learning models, particularly the black-box models, are widely favored for their outstanding predictive capabilities. However, they often face scrutiny and criticism due to the lack of interpretability. Paradoxically, their strong predictive capabilities may indicate a deep understanding of the underlying data, implying significant potential for interpretation. Leveraging the emerging concept of knowledge distillation, we introduce the method of knowledge distillation decision tree (KDDT). This method enables the distillation of knowledge about the data from a black-box model into a decision tree, thereby facilitating the interpretation of the black-box model. Essential attributes for a good interpretable model include simplicity, stability, and predictivity. The primary challenge of constructing interpretable tree lies in ensuring structural stability under the randomness of the training data. KDDT is developed with the theoretical foundations demonstrating that structure stability can be achieved under mild assumptions. Furthermore, we propose the hybrid KDDT to achieve both simplicity and predictivity.  
 An efficient algorithm is provided for constructing the hybrid KDDT. Simulation studies and a real-data analysis validate the hybrid KDDT’s capability to deliver accurate and reliable interpretations. KDDT is an excellent interpretable model with great potential for practical applications.
	\end{abstract}
			
	\noindent%

	\spacingset{1.5} 

\section{Introduction}\label{sec:intro}

In recent decades, machine learning (ML) has gained significant popularity in various fields. However, the widespread adoption of black-box ML models, such as neural networks and ensemble models, has led to growing concerns about their interpretability. This lack of interpretability has triggered skepticism and criticism, particularly in decision-based applications. For instance, in fields like medical diagnostics or treatment choice, without straightforward and concise interpretability the model could lead to erroneous diagnoses and potentially harmful treatment decisions. Knowledge distillation \citep{BaNIPS2014,hinton2015distilling,urban2017deep,shi2019knowledge,stanton2021does,allenzhu2021understanding} provides a way to interpret the black-box ML model through a transparent model, following a teacher-student architecture \citep{Hu2023}. Knowledge about the data is distilled from the teacher model (black-box ML model) to train the student model (transparent model). As a result, the student model (transparent model) inherits the teacher model’s knowledge about the complex structure of the data and the underlying mechanisms of the domain question, enabling it to achieve both high interpretability and strong predictive performance.
\cite{Ribeiro2016} employed several simple models, including linear models and decision trees, as transparent models. Similarly, \cite{Lundberg2017} utilized kernel methods and local linear approaches to construct the transparent model. 
In this study, we focus on the decision tree \citep{Johansson2011,frosst2017distilling,Coppens2019DistillingDR,e22111203,Song_2021_CVPR,ding2021cdt} which emerges as an ideal transparent model for two reasons. First, it is inherently interpretable. Second, it possesses the capacity to capture complex data structures. 
Several studies have employed decision tree as transparent model alongside knowledge distillation. \cite{frosst2017distilling} explored the distillation of a neural network into a soft decision tree. \cite{Coppens2019DistillingDR} discussed the decision tree model in interpreting deep reinforcement model. \cite{ShenXC20} used decision tree for explaining data in the field of e-commerce. However, none of these studies considered the stability of decision trees constructed through knowledge distillation. 
The interpretability of decision tree relies heavily on the stability of its structure, which may be sensitive to the specific datasets used for training. Interpretations may become questionable if minor changes in the training data significantly affect the tree's structure. Given that the training data is generated randomly through the knowledge distillation process, ensuring the stability of the tree's structure becomes a key challenge to address.
\cite{zhou2018approximation} explored tree structure stability in knowledge distillation, while their study focused on a single splitting criterion and did not provide conclusive conditions under which the tree structure (or split) converge.

We refer to the decision tree generated from the knowledge distillation process as the ``knowledge distillation decision tree'' (KDDT). In this paper, we conduct a comprehensive theoretical study for the split stability of KDDT, demonstrating that split will converge in probability with a specific convergence rate, subject to mild assumptions. Our theoretical findings encompass the most commonly used splitting criteria and are applicable to both classification and regression applications. Additionally, we propose and implement algorithms for KDDT induction. 
Note that KDDT provides a global approximation to the black-box model, meaning it approximates the entire black-box model at once using a single interpretable model. This approach may be less efficient for interpreting very large and complex black-box models, such as deep neural networks, compared to the local approximation models discussed in \cite{Zhou2016}, which approximate the black-box model piecewisely. For local approximation model, each segment of the black-box model can be represented by a different local approximation, allowing for more tailored interpretations.
However, large-scale black-box models have large number (e.g., millions) of parameters and require large training datasets, making them not suitable for small or medium datasets, such as those with fewer than or equal to $O(10^3)$ samples. For these datasets, the global approximation provided by KDDT can offer more accurate interpretations for the global effects of covariates than simple linear models. Through a simulation study, we validate KDDT's ability to provide precise interpretations while maintaining a stable structure. We also include real data analysis to demonstrate its practical applicability.

The rest of the paper is organized as follows. In Section \ref{sec:MDDT}, we introduce the concept and stability theory of KDDT. The algorithms for constructing KDDT are proposed in Section \ref{sec:ALG}. Section \ref{sec:Sim}  presents the simulation study. In Section \ref{sec:apply}, we apply KDDT on real datasets. Finally, we conclude and engage in a discussion in Section \ref{sec:Dis}. Theorems and proofs are in Appendix \ref{apdxA}. Supplementary materials can be found in Appendix \ref{apdxB}. Additionally, an open-source R implementation of KDDT is accessible on GitHub at https://github.com/lxtpvt/kddt.git.

\section{Knowledge Distillation Decision Tree}\label{sec:MDDT}
A knowledge distillation decision tree is essentially a decision tree. Instead of being constructed from real observations, it is generated from the knowledge distillation process.

\subsection{Knowledge Distillation Process}\label{TIvKD}
\begin{figure*}[ht]
\centering
\includegraphics[width=.9\textwidth]{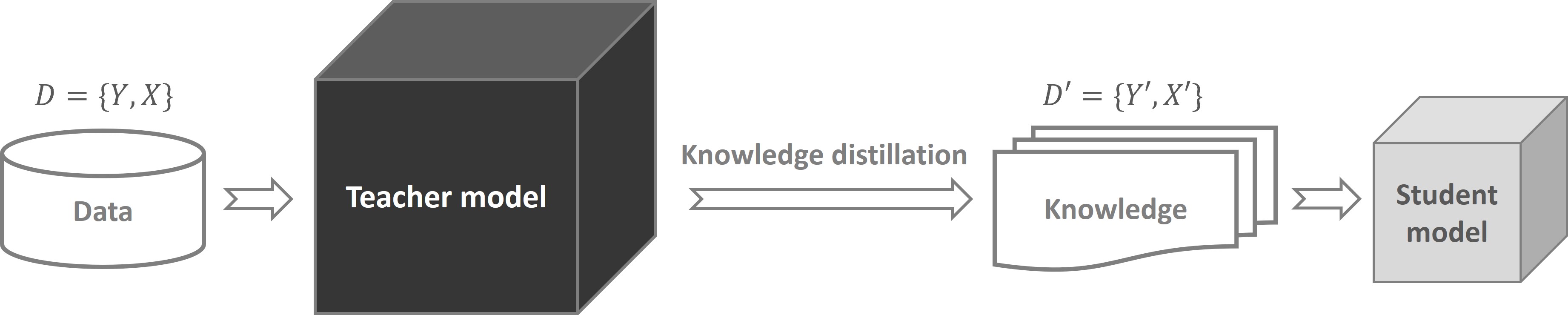}
\caption{\small Teacher-student architecture for knowledge distillation.}
\label{fig:KDP}
\end{figure*}

A typical knowledge distillation process with the teacher-student architecture is illustrated in Figure \ref{fig:KDP}. The specific components of this process can be adapted based on application requirements. For example, the teacher model can be a Convolutional Neural Network (CNN) \citep{Gou2021IJCV, Kaleem2024ACR} or a Large Language Model (LLM) \citep{Yang2024}, while the student model can be a decision tree \citep{Johansson2011, frosst2017distilling, Coppens2019DistillingDR} or a lightweight neural network \citep{Gou2021IJCV, Kaleem2024ACR}. In this paper, we specify the components as follows:

\begin{itemize}
   \item \textbf{Data.} $D=\{Y,X\}$, where $Y$ is the set of observations of response variable $y$, $X$ is the set of observations of covariates $\mathbf{x}=( x_1,...,x_p)$. Both response and covariates can be categorical or continuous variables.
   \item \textbf{Teacher model.} $y = f(\mathbf{x})$, we specify it as a small scale black-box ML model for the size of data $O(10^3)$.
  \item \textbf{Knowledge distillation.} Includes two steps: 1) random sampling of covariate values on their support, denoted as $X'$, and 2) generating the corresponding response values $Y'=f(X')$ through the fitted teacher model $f$.
   \item \textbf{Knowledge.} $D' = \{Y', X'\}$, we call it pseudo-data.
   \item \textbf{Student model.} A decision tree, we refer to it as a knowledge distillation decision tree (KDDT). The KDDT is constructed from the pseudo-data $D'$ and keeps a stable structure under the randomness of $D'$.
\end{itemize}

The knowledge distillation process can also be viewed as a model approximation process, as illustrated in Figure \ref{fig:approximation} in Appendix \ref{apdxB}. The student model, KDDT, is used to approximate the teacher model through the pseudo-data $D' = \{Y', X'\}$. Additionally, Figure \ref{fig:approximation} also highlights the differences between the model approximation and generalization processes.

\subsection{Tree Structure Stability}
\label{sec:TSS}
In this paper, we focus on the second half of the knowledge distillation process, specifically from knowledge distillation to student model. The teacher model and original data are fixed. Our task is to handle the randomness in the pseudo-data $D'$ and construct a stable KDDT. It is based on the hypothesis that we can generate arbitrarily large $D'$ to achieve the stability of KDDT. In this section, we will prove this hypothesis.

\subsubsection{Prerequisites}\label{sec:prereq}
It is essential to introduce the key concepts and notations of decision tree that will be used in the theoretical study.

\noindent{(1) Splitting criteria}

Typically, different criteria are used for regression and classification trees. 
In regression, the primary criteria include minimizing the sum of squared errors (SSE) or the mean squared error (MSE) after splitting:
\begin{equation}\label{ssr}
\begin{split}
& min \{\sum^{n_l}_{i=1}(y_{li}-\Bar{y}_l)^2+\sum^{n_r}_{j=1}(y_{rj}-\Bar{y}_r)^2 \}, \\
& min \{ \frac{1}{n_l}\sum^{n_l}_{i=1}(y_{li}-\Bar{y}_l)^2+\frac{1}{n_r}\sum^{n_r}_{j=1}(y_{rj}-\Bar{y}_r)^2 \},
\end{split}
\end{equation}
where the subscripts $l,r$ represent the left and right node of a stump, $n_l+n_r=n$, $\Bar{y}_l = \frac{1}{n_l}\sum^{n_l}_{i=1}y_{li}$ and $\Bar{y}_r = \frac{1}{n_r}\sum^{n_r}_{j=1}y_{rj}$.

In classification, the criterion for selecting the best split is to maximize the reduction of impurity after splitting:
\begin{equation*}
   max \{ E-(E_l+E_r) \},
\end{equation*}
where $E$ is the total impurity before splitting, and $E_l$ and $E_r$ are the left and right child impurities, respectively, after splitting. Since the split does not impact $E$, above criterion can be simplified as follows:
\begin{equation}
  \label{entropy}
   min \{ E_l+E_r \},
\end{equation}
The well-known impurity measures include Shannon entropy, gain ratio, and Gini index \citep{quinlan1993c45, BreiFrieStonOlsh84}. \cite{Wang2017ICASSP} proposed the Tsallis entropy in \eqref{tsallis} to unify these measures in a single framework.
\begin{equation}
  \label{tsallis}
  E=S_q(Y)=\frac{1}{1-q}(\sum_{i=1}^{C}p(y_i)^q-1), \hspace{5mm} q \in \mathbb{R},
\end{equation}
where $Y$ is a random variable that takes value in $\{y_1,...,y_C\}$, $p(y_i)$ is the corresponding probabilities of $y_i$, $i=1,...,C$, and $q$ is an adjustable parameter.

\noindent{(2) Split search algorithm}

The most commonly used split search algorithm is the greedy search algorithm, which makes a locally optimal choice at each stage in a heuristic manner, to find a global optimum. The algorithm involves the steps: (a) for each split, searching through all covariates and their observed values; (b) for each candidate pair (covariate, value), calculating the loss (gain) defined by splitting criterion; and (c) identifying the best split by minimizing the loss (maximizing the gain). Although the greedy search algorithm may not guarantee the global optimum which is theoretically an NP problem \citep{HYAFIL1976}, we still choose it for our study due to its simplicity and popularity in practice.


\subsubsection{Split convergence}\label{SR}
As discussed in the introduction, studying the stability of the entire tree is challenging. A practical approach is to focus on individual splits. If all splits are stable, the entire tree is stable naturally.
We refer to a split as achieving stability when it converges to a unique optimal split. The concepts of optimal split is defined as follows. 
\begin{defin}[Optimal split]
\label{df-os}
Let $\Omega$ be the support of univariate $x$, and $z^l_i(x)$ and $z^r_i(x)$ are functions $\Omega \to \mathbb{R}$, where $i=1,...,C$ and $C$ is a constant in $\mathbb{N^+}$. Let $g(z^{t}_1(x),...,z^{t}_C(x))$ be a continuous function $\mathbb{R^C}\to \mathbb{R}$, where $t=l$ or $r$.
Then, the optimal split $x_s\in\Omega$ is defined as follows.
\begin{equation}
  \label{def-1}
  x_s=\underset{x\in\Omega}{\operatorname{argmin}}[g(z^l_1(x),...,z^l_C(x))+g(z^r_1(x),...,z^r_C(x))].
\end{equation}
\end{defin}

Definition \ref{df-os} is somewhat abstract. To clarify, we provide two examples to illustrate this definition in both regression and classification contexts.
\begin{itemize}
    \item Regression: we assume that both $y$ and $x$ are continuous variables, and that the split criterion is the MSE as defined in \eqref{ssr}. The components in Definition \ref{df-os} are outlined as follows.
    $$g(z^l_1(x))=z^l_1(x), \hspace{2mm} g(z^r_1(x))=z^r_1(x),$$
    $$z^l_1(x) = \int^x_a (f(t)-\mu_l(x))^2 \,dt, $$
    $$ z^r_1(x) = \int^b_x (f(t)-\mu_r(x))^2 \,dt,$$
    where,
$$ \mu_l{(x)}= \frac{1}{x-a} \int_{a}^{x} f(u) \,du , \hspace{2mm} \mu_r{(x)}=\frac{1}{b-x} \int_{x}^{b} f(u) \,du. $$

    \item Classification: we assume that $y \in \{y_1,...,y_C\}$ is a categorical variable with $C$ categories, $x$ is a continuous variable, and the split criterion is defined by \eqref{entropy} using the Tsallis entropy as in \eqref{tsallis}. The components in Definition \ref{df-os} are specified as follows.
    $$g(z^l_1(x),...,z^l_C(x))=\frac{1}{1-q}(\sum_{i=1}^{C}z^l_i(x)^q-1),$$
$$g(z^r_1(x),...,z^r_C(x))=\frac{1}{1-q}(\sum_{i=1}^{C}z^r_i(x)^q-1),$$
\begin{equation*}
        z^l_i(x) = \int^x_a \frac{1}{x-a} * I_{y_i}(f(t)) \,dt ,
\end{equation*}

\begin{equation*}
        z^r_i(x) =\int^b_x \frac{1}{b-x} * I_{y_i}(f(t)) \,dt ,
\end{equation*}
where $I_{y_i}(f(x))$ is an indicator function that is equal to 1 at $f(x)=y_i$ and 0 elsewhere.
\end{itemize}

The concept of split convergence can be defined based on the definition of an optimal split as follows.
\begin{defin}[Split convergence]
\label{split_convergence}
A split $x^n_s$ is estimated via greedy search algorithm on the sampled data $D'=\{Y', X'\}$ with size $n$. Let $x_s$ be the unique optimal split on $\Omega$. If $x^n_s$ converges to $x_s$ in probability as $n \to \infty$, we refer to this case as split convergence and $x^n_s$ as a convergent split.
\end{defin}

Our theoretical study demonstrates that split convergence can be guaranteed under three assumptions: (1) the existence of unique optimal split; (2) the uniform random sampling of pseudo-data $D'=\{Y', X'\}$; and (3) the greedy search algorithm. Since continuous and categorical response variables have different split criteria for regression and classification, and different types of covariates require distinct treatments in the proof, we divide the theory into four theorems, each corresponding to one of the combinations of variable types listed in Table \ref{table-1}. The details of all theorems, lemma, and their proofs can be found in Appendix \ref{apdxA}.

\begin{table}[t!]
\small
\caption{\small Theorems classified based on the combinations of variable types.}
\centering
\begin{tabular}{|l||*{3}{c|}}\hline
\backslashbox{$x \hspace{0.7cm}$}{$y$}
&\makebox[3 em]{Continuous}&\makebox[3 em]{Categorical}\\\hline\hline
Continuous & Theorem \ref{th:scenario1} & Theorem \ref{th:scenario2}\\\hline
Categorical & Theorem \ref{th:scenario3} & Theorem \ref{th:scenario4}\\\hline
\end{tabular}
\label{table-1}
\end{table}

Fair assumptions help establish a theory with a solid foundation and broad applicability. Regarding the greedy search assumption, as discussed in Section \ref{sec:prereq}, it has the advantages of simplicity and popularity in practice. The uniform random sampling assumption ensures the sampling space covers the teacher model and simplifies theoretical proofs. However, it may lead to efficiency issues as the dimension of covariates increases. For problems with modest dimensions (the number of continuous variables $<20$), uniform random sampling works well (see real data analysis in Section \ref{sec:apply}). For high-dimensional problems, non-uniform random sampling strategies may be more appealing and worth investigating. As for the unique optimal split assumption, let's consider its opposite first: assume there are multiple optimal splits. We can define the concept of split oscillation in Definition \ref{df_so}. Although split oscillation may occur in theory, it rarely happens in practice. Let's consider a scenario where two optimal splits exist. When applying the greedy search algorithm with real data, the likelihood of two splits yielding identical numerical results (e.g., impurity reduction) will be extremely low. Even if such a rare situation arises, it's not a significant concern. It simply indicates that the two splits are equivalent, and selecting either of them is reasonable.

\begin{defin}[Split oscillation]
\label{df_so}
A split $x^n_s$ is estimated via greedy search algorithm on the sampled data $D'=\{Y', X'\}$ with size $n$. If $x^n_s$ has multiple limits as $n \to \infty$, we refer to this case as split oscillation and the split as an oscillating split.
\end{defin}

\subsubsection{Measure of split stability}\label{sec:MoS}
\begin{figure}[ht]
\centering
\includegraphics[width=0.7\textwidth]{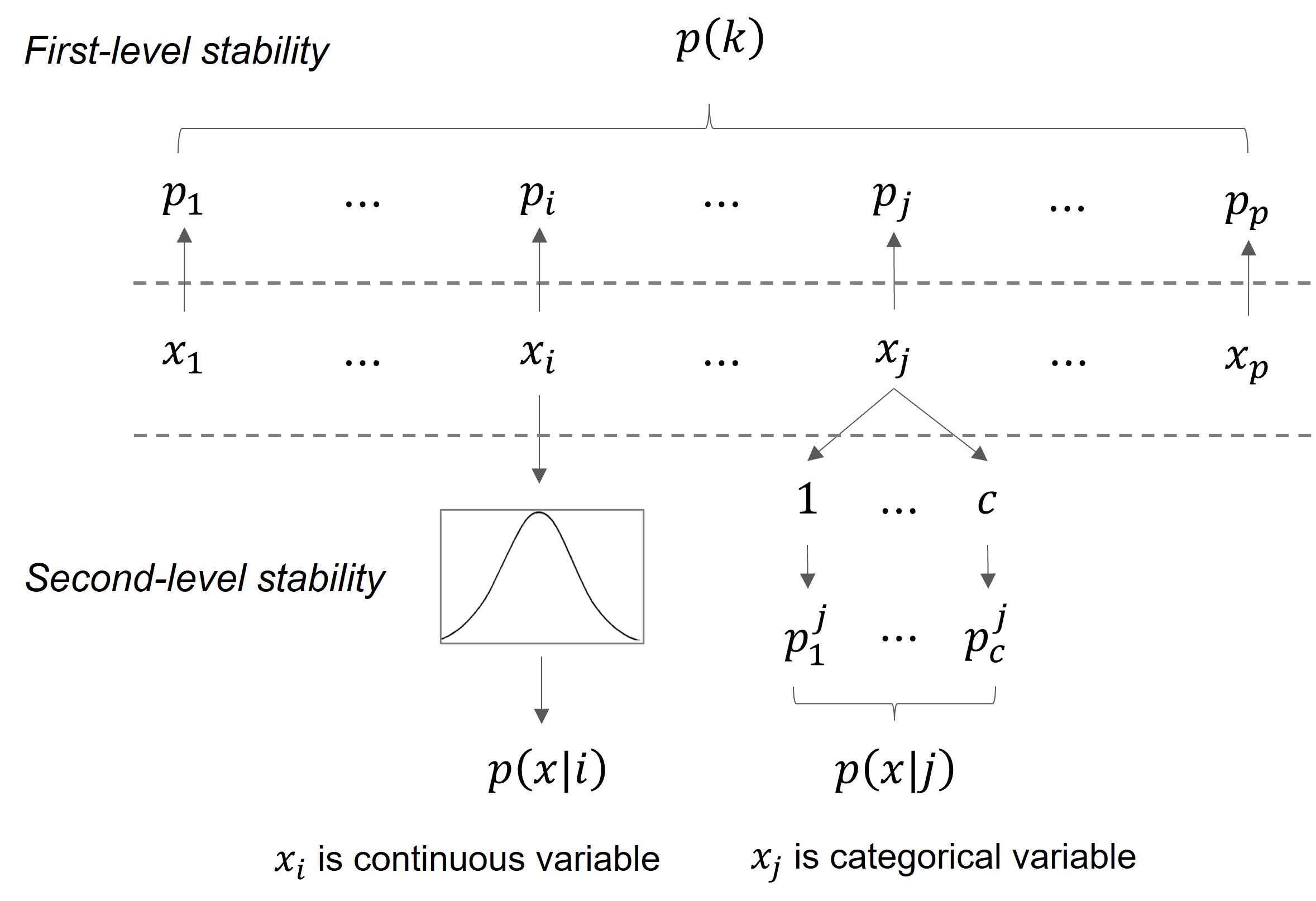}
\caption{\small Two-level split stability. First-level stability is denoted as the pmf of choosing a split variable. Second-level stability is denoted as either a pdf or a pmf conditioning on the selected split variable.}
\label{fig:2LevelStability}
\end{figure}

In practice, the pseudo-data must be finite. Therefore, we need a way to measure the split stability under finite data. Motivated by the greedy search algorithm, we propose the two-level split stability (see Figure \ref{fig:2LevelStability}) as follows.

\begin{itemize}
   \item \textbf{First-level stability.}
    It is defined as a discrete distribution with the probability mass function $p(k)$, which quantifies the stability of selecting the $k$-th covariate $x_k$, ($k = 1,...,p$) as the splitting variable.
    
   \item \textbf{Second-level stability.} 
    It is defined as the conditional distribution of the splitting value given the covariate to split on. The second-level stability can be either a probability density function (e.g., $p(x|i)$) or a probability mass function (e.g., $p(x|j)$), depending on the type of selected splitting variable.
\end{itemize}

Since the two-level stability is difficult to calculate analytically, we use Monte Carlo simulation for its estimation.


\section{Algorithms for Constructing KDDT}
\label{sec:ALG}
There are two fundamental distinctions in the construction of KDDT compared to ordinary decision tree (ODT). Firstly, ODT is built directly from a limited dataset, whereas KDDT is constructed using unlimited (in theory) pseudo-data. Secondly, ODT's goal is to best fit the dataset, whereas KDDT's objective is to best approximate the teacher model. These distinctions result in a different construction algorithm of KDDT compared to ODT.

\subsection{KDDT Induction Algorithm}\label{sec:TreeIAs}
We first introduce the concept of the sampling region, which will be utilized in the KDDT induction algorithm.
\begin{defin}[Sampling region]\label{def:SR}
Let $\mathbb{S}$ be the bounded space defined by the observed data. For node $i$ in KDDT, its ancestors define a subspace on $\mathbb{S}$. We denote it as $R_i$ and refer to it as the sampling region of node $i$. The sampling region of any inner node is exactly the union of the sampling regions of its two child nodes. (Note: Since the boundary of observed data is limited, $\mathbb{S}$ is bounded.)
\end{defin}

\begin{figure*}[ht]
\centering
\includegraphics[width=\textwidth]{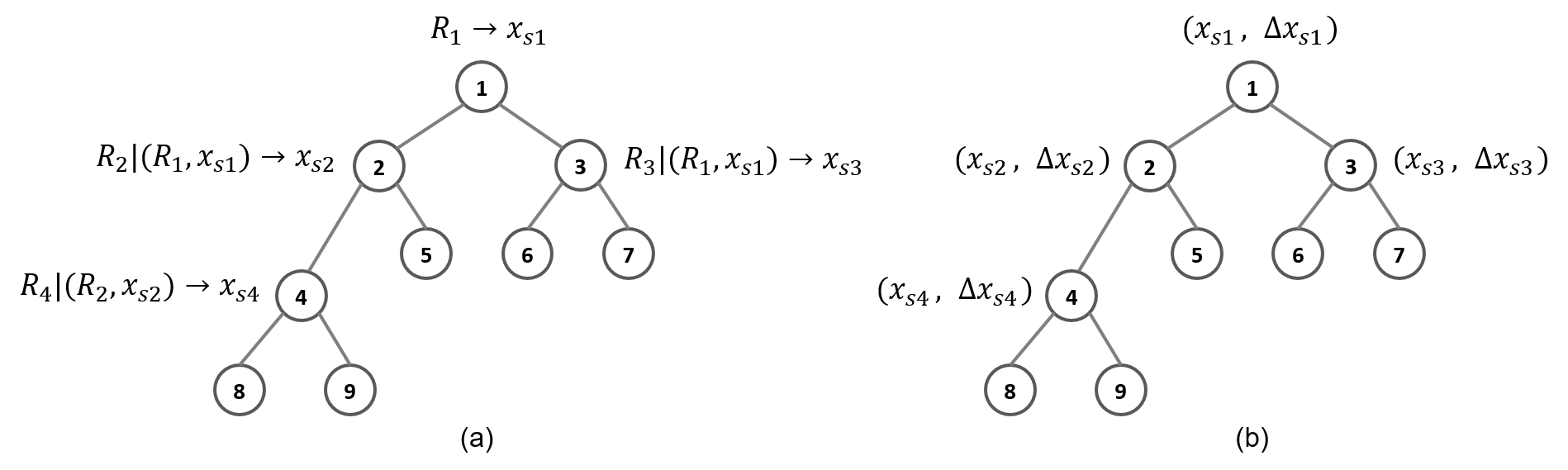}
\caption{\small Examples of the dependency chain and variance propagation. (a) Two dependency chains in a tree. (b) The corresponding variance propagation. Note: $R_i$ denotes sampling region $i$. $R_2|(R_1,x_{s1}) \to x_{s2}$ indicates that $x_{s2}$ is determined by $R_2$, given $R_1$ and $x_{s1}$. The variance of $x_{si}$ is denoted by $\Delta x_{si}$.}
\label{ddt-8}
\end{figure*}

As an extension of the sampling region, the concept of the sampling path will also be used later in this paper.
\begin{defin}[Sampling path]
\label{def:spath}
A sampling path is a series of nested sampling regions defined by the nodes in a KDDT path. The sampling path $P_{i,j}$ starts from sampling region $R_i$ and ends at sampling region $R_j$, i.e., $P_{i,j}=\{(R_{i},...,R_{j}) | R_{i}\supset...\supset R_{j}\}$. Two sampling paths intersect if there exists a sampling region in one sampling path that includes any sampling region in the other sampling path.
\end{defin}
The most commonly used induction algorithm for constructing an ODT is a top-down recursive approach \citep{Rokach2014book}, referred to as the ODT induction algorithm in this paper. It starts with the entire input dataset in the root node, where a locally optimal split is identified using the greedy search algorithm, and conditional branches based on the split are created. This process is repeated in the generated nodes until the stopping criterion are met. 
A naive approach to constructing a KDDT is to directly apply the ODT induction algorithm on a large pseudo-dataset. However, this method may not perform well in practice. The pseudo-data introduces variation (uncertainty) due to the random sampling process. This variation propagates in the constructed tree along dependency chains created by the top-down induction strategy. For instance, as illustrated in Figure \ref{ddt-8} (a), the split $x_{s4}$ depends on $(R_2,x_{s2})$, which, in turn, depends on $(R_1,x_{s1})$. The propagation of split variance follows the inverse direction of these dependencies. We denote the variance of $x_{si}$ as $\Delta x_{si}$. In Figure \ref{ddt-8} (b), the variance $\Delta x_{s1}$ will affect $(x_{s2},\Delta x_{s2})$ and $(x_{s3},\Delta x_{s3})$, and subsequently, $\Delta x_{s2}$ will impact $(x_{s4},\Delta x_{s4})$. This results in rapid inflation of variance as it propagates to deeper levels. For example, a small $\Delta x_{s1}$ may lead to a substantial $\Delta x_{s4}$ or even a change in the split variable. 

Incorporating the two-level stability, we proposed a KDDT induction algorithm to avoid the variation inflation issue in the ODT algorithm. As shown in the steps of KDDT induction algorithm, for a given node $i$, we measure its split $N_i$ times. Utilizing these repeated measurements represented as $X_s$, we can calculate the two-level stability and choose a split value with the lowest variance. The first-level stability aids in reducing the variance when selecting the split variable, while the second-level stability assists in reducing the variance when identifying the split value. For example, if the split variable is continuous, considering the mean of all fitted split values $\bar{x}_s$, by central limit theorem, the variance of $\bar{x}_s$ will reduce at a rate of $N^{-1}_i$. Instead of using the mean of all fitted split values, the two-level stability approach choose the mode, which not only reduces variance but also mitigates the influence of outliers. Additionally, choosing the mode aligns with the likelihood principle, as it corresponds to selecting the value with the maximal likelihood, given that two-level stability is defined by probability mass/density function. By repeating this process at each split, we can construct a KDDT with a stable structure. In practice, it is common to choose a reasonably large value for $N_i$ ($N_i=100$ works well for our study).
The sample size of the pseudo data $n_i$ can be estimated by using (proportional to) the potential explanation index (see Definition \ref{def:XI}). In practice, if the number of nodes is small (see interpretable nodes in Section \ref{sec:hybridDDT}), we can simply set all $n_i$ to be the same at the same tree level and assign their values equal to 90\% of the corresponding value in the preceding upper level. We select 90\% to maintain a large number of pseudo-data, ensuring a stable estimation of the split value.
We determine the value of $x^*_s$ by selecting the mode of the second-level stability (pmf/pdf). The stopping criterion can be defined as the ratio of prediction accuracy (e.g., MSE or C-Index) between the teacher model and KDDT on the observed data, evaluated through cross-validation.

\RestyleAlgo{ruled}
\begin{algorithm}[t!]\small
\label{alg:ddtInduction}
\caption{Steps of KDDT induction algorithm}
\vspace{3mm}
\KwData{pseudo-data}
\KwResult{A knowledge distillation decision tree}
Starting from the root node, set $i=1$, and create an empty set $X_s$ to store splits.

\While{ the stopping criterion is not met,}{
\vspace{2mm}
\begin{enumerate}
   \item For node $i$, repeat the following processes $N_i$ times.
   \begin{enumerate}
       \item [(1)] Generate pseudo-data, which includes $n_i$  \\  samples from the sampling region $R_i$  \\  corresponding to node $i$.
       \item [(2)] Fit a stump on the pseudo-data and store the  \\  split of the stump into $X_s$.
   \end{enumerate}
   \item Compute two-level stability with $X_s$ to identify the  \\  best split $x^*_s$ (the mode of the second-level stability).
   \item Apply $x^*_s$ to create child nodes and set their id as  \\  $2i$, $2i+1$, respectively.
   \item Move to the next node that needs to be split.
\end{enumerate}
}
\end{algorithm}

\subsection{Hybrid Induction Algorithm and Hybrid KDDT}\label{sec:hybridDDT}

We assume that the teacher model is well-defined, meaning it is well-fitted to the observed data without overfitting. Since KDDT aims to optimize its approximation to the teacher model, we can ignore any overfitting concerns for KDDT in relation to the observed data. Thus, we can focus solely on achieving a balance between the degree of approximation and computational efficiency during KDDT construction.
KDDT induction algorithm requires repetitive sampling and fitting, performed $N_i$ times to identify the best split. Each time, the pseudo-data need to have a sufficient size, leading to high computational load. Furthermore, to achieve a high-quality approximation, the tree needs to grow to a large size. Consequently, growing a large tree solely using KDDT induction algorithm is often computationally infeasible.

\begin{figure}[t!]
\centering
\includegraphics[width=.35\textwidth]{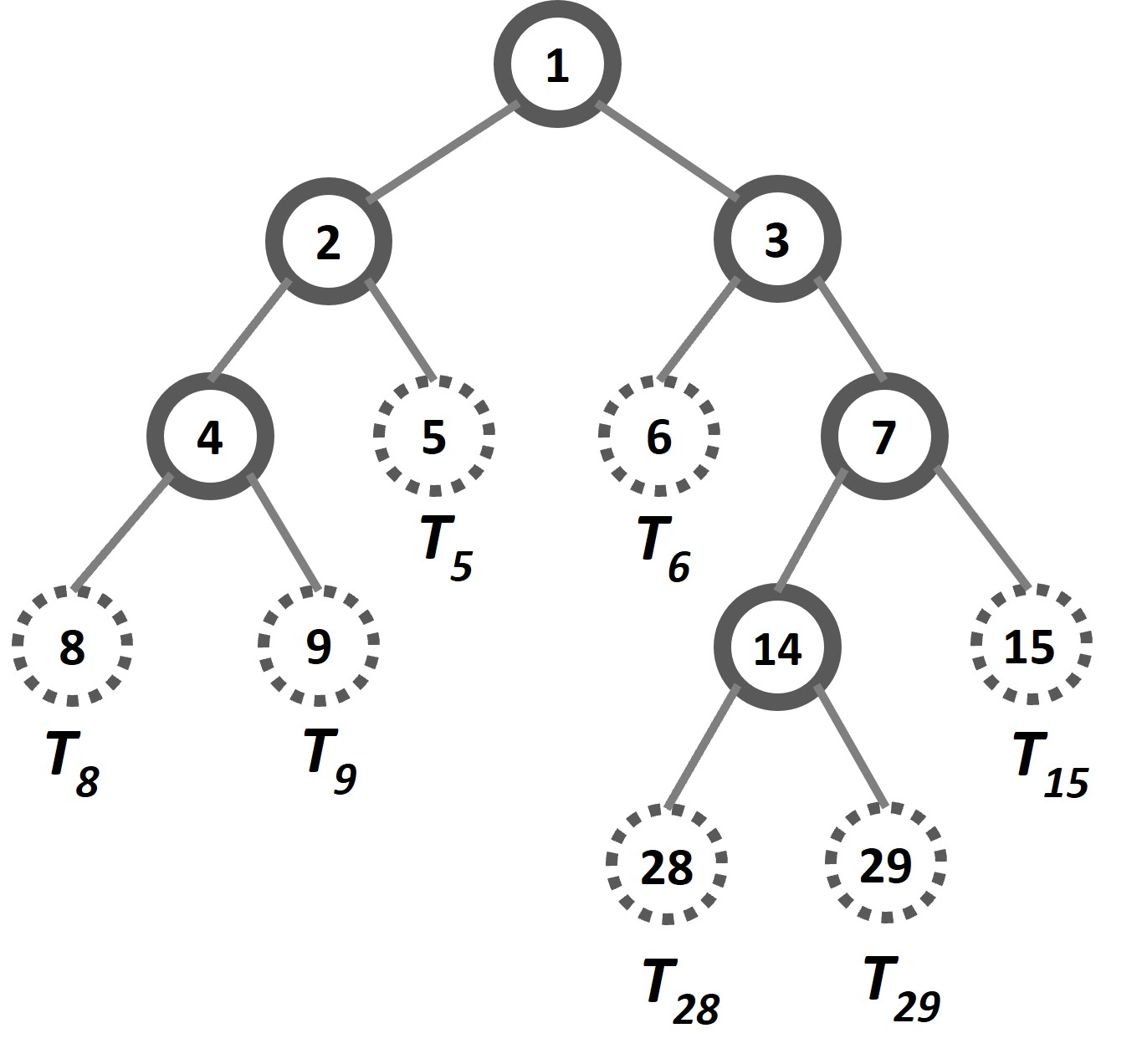}
\caption{\small An example of the hybrid knowledge distillation decision tree: the nodes $\{1, 2, 3, 4, 7, 14\}$ are interpretable nodes, while all other nodes are predictive nodes.}
\label{fig:HybridDDTSampleStrategy}
\end{figure}

Typically, in most real-world applications, only a small set of splits is needed for interpretation purposes. We refer to these splits as interpretable nodes (splits), while all other nodes, i.e., terminal nodes, are named predictive nodes. Since the ODT induction algorithm is much more efficient in constructing large trees than KDDT's, it is reasonable to combine these two algorithms to a hybrid induction algorithm. Specifically, we apply the KDDT induction algorithm to the interpretable nodes, ensuring their two-level stability, which is crucial for interpretation. Then, we employ the ODT induction algorithm to construct large sub-trees at the predictive nodes, maintaining a good approximation to the teacher model and increasing the computation efficiency. We refer to this tree as a hybrid KDDT. 
For instance, in Figure \ref{fig:HybridDDTSampleStrategy}, the interpretable nodes are $\{1, 2, 3, 4, 7, 14\}$. We use the KDDT induction algorithm to identify the stable splits for these nodes. Then, we employ the ODT induction algorithm to grow the large sub-trees $\{T_5,T_6,T_8,T_9,T_{15},T_{28},T_{29}\}$ at the respective predictive nodes.

The small set of interpretable nodes enhances the simplicity of model interpretation. Meanwhile, the complexity necessary to ensure prediction accuracy comparable to that of the teacher model is achieved through the construction of large sub-trees at the predictive nodes. This decoupling between interpretability and complexity offers the potential for hybrid KDDT to strike a balance between prediction accuracy and interpretability. For the sake of simplicity, we use the term ``KDDT'' to refer to hybrid KDDT in the remainder of this paper.

The informativeness of the interpretation can vary across different interpretable nodes. Measuring and reporting these differences is crucial for the interpretation according to these nodes. To address this issue, we introduce the concept of explanation index (XI). A similar index is calculated and referred to as the potential explanation index (PXI) for the predictive nodes.

\begin{defin}[Explanation Index and Potential Explanation Index]
\label{def:XI}
The explanation index of interpretable node (split) $i$, denoted as $XI_i$, and the potential explanation index of predictive node $j$, denoted as $PXI_j$, are defined as follows:
\begin{equation}\label{eq:XI}
XI_i = \frac{\frac{n_{i}}{n}*\Delta_{S_i}}{\Delta_{KDDT}} * 100\%, \hspace{3mm}
PXI_j = \frac{\frac{n_{j}}{n} * \Delta_{T_j}}{\Delta_{KDDT}}*100\%
\end{equation}
where $n_i$, $n_j$, and $n$ denote the number of observations in node $i$, $j$, and the entire dataset. $\Delta_{S_i}$ and $\Delta_{T_j}$ represent the change in the measure defined by the split criterion (e.g., impurity or MSE reduction) after fitting split $i$ or the subtree at node $j$, respectively. $\Delta_{KDDT}=\sum_i \frac{n_{i}}{n}*\Delta_{S_i} + \sum_j \frac{n_{j}}{n} * \Delta_{T_j}$. 
\end{defin}

Based on Definition \ref{def:XI}, it is straightforward to verify that $\sum_iXI_i+\sum_jPXI_j=1$. $XI_i$ and $PXI_j$ can be considered as information contained in the interpretable node $i$ and predictive node $j$, respectively.
Furthermore, we can extend the concept of XI to apply to a path in KDDT as follows.
\begin{defin}[Path Explanation Index]
\label{def:XIP}
\begin{equation}\label{eq:XIP}
XI_{ij} = \sum_{k\in S_{ij}} XI_k
\end{equation}
where node $i$ is an interpretable node, node $j$ is a descendant of node $i$, and $S_{ij}$ is a set of node IDs that includes the nodes in the path from node $i$ to the parent of node $j$.
\end{defin}
With above indices, we can identify the desired hybrid KDDT with an appropriate number of interpretable nodes. For instance, if we want to achieve more than 70\% of the information in the data explained by the interpretable nodes, the stopping criterion is $\sum_iXI_i > 70\%$, i.e., $\sum_jPXI_j < 30\%$. Examples demonstrating their applications can be found in Section \ref{sec:apply}. The process of constructing a desired hybrid KDDT with appropriate number of interpretable nodes is illustrated by an example shown in the Figure \ref{fig:findHybridDDT} in Appendix \ref{apdxB}. The potential explanation index of a predictive node can also be used to determine the size of the pseudo data in its sampling region which could be proportional to its PXI. Because, a higher PXI indicates greater unexplained information, requiring a larger pseudo data size.

\section{Simulation Study}\label{sec:Sim}
The simulation study has three primary objectives: (1) to demonstrate the effectiveness of KDDT in revealing intricate structures of the data, (2) to validate the interpretability of KDDT, and (3) to illustrate the stability of interpretable splits (nodes). 

\begin{figure*}[t!]
\centering
\includegraphics[width=\textwidth]{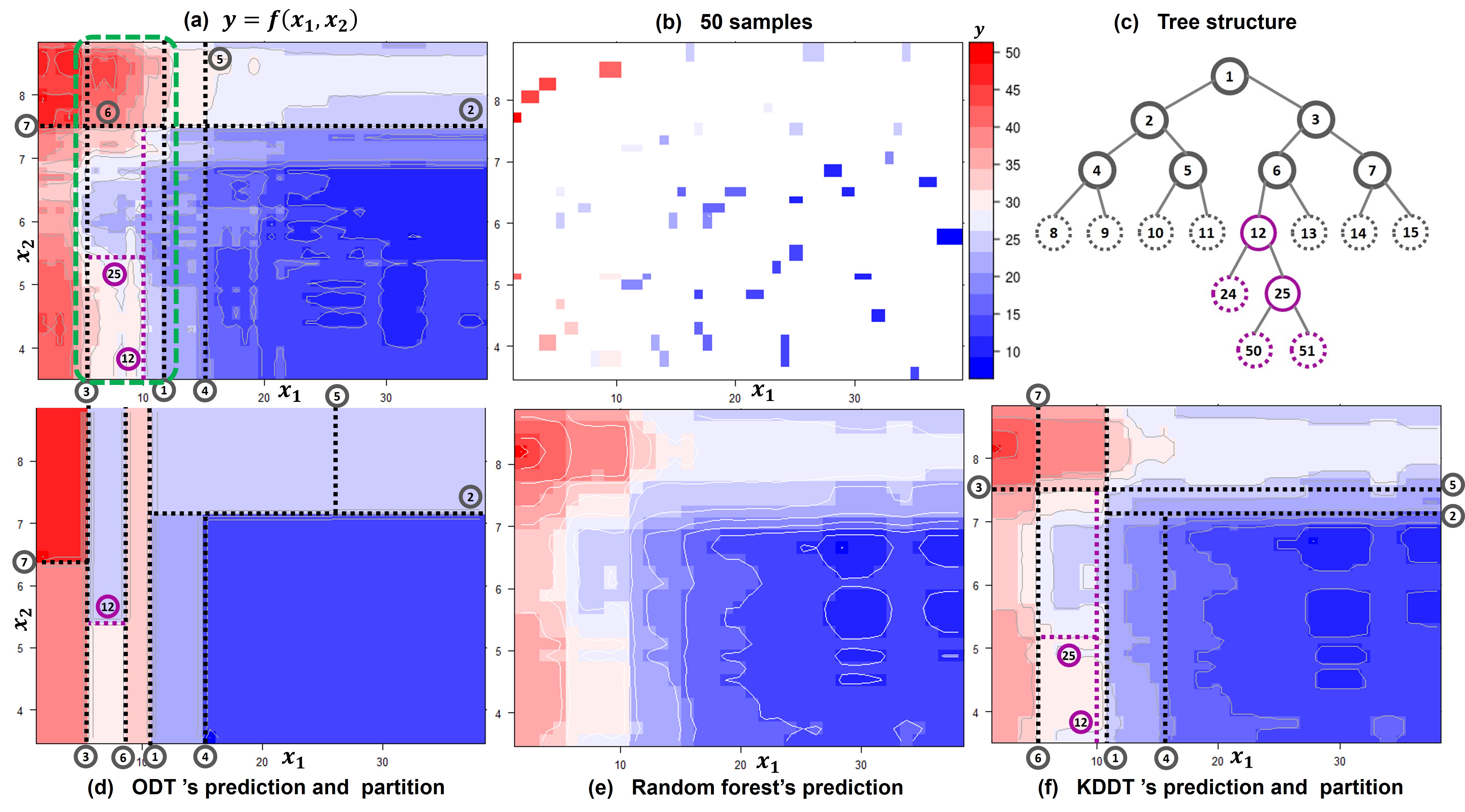}
\caption{\small The effectiveness of KDDT in revealing and explaining complex data structures. (a) The true function $y=f(x_1,x_2)$ and its partition with 9 splits. (b) 50 random samples from the true function. (c) The tree structure and splits for defining the partitions in (a), (d), and (f). (d) ODT is fitted based on the 50 samples. (e) RF is fitted based on the 50 samples. (f) KDDT is constructed based on the RF.}
\label{fig:sim2D}
\end{figure*}

To facilitate a clear and intuitive discussion, we introduce a two-dimensional function $y=f(x_1,x_2)$, consisting of 2601 generated data points, as illustrated in panel (a) of Figure \ref{fig:sim2D}. This function exhibits high non-linearity and intricate interactions, making it well-suited for our purposes. Let's assume that $y=f(x_1,x_2)$ is unknown. We can gain insights about it by analyzing the sampled observations. In panel (b), we have 50 observations randomly sampled from the true function. We want to compare KDDT with other interpretable models. The well-known ones include linear regression and decision tree (ODT). As linear regression is not suitable for this data, we opt for ODT. The ODT fitted from 50 observations is presented in panel (d). In comparison to the true function, the ODT estimation is coarse and unable to capture the interaction structure within the area marked by the green rectangle. In contrast, the random forest model provides a refined and precise estimation, as shown in panel (e). The KDDT presented in panel (f), as a close approximation of its teacher, maintains a high-quality (resolution) estimation. This highlights the ability of the KDDT to reveal intricate structures in the data.

\begin{figure*}[t!]
\centering
\includegraphics[width=\textwidth]{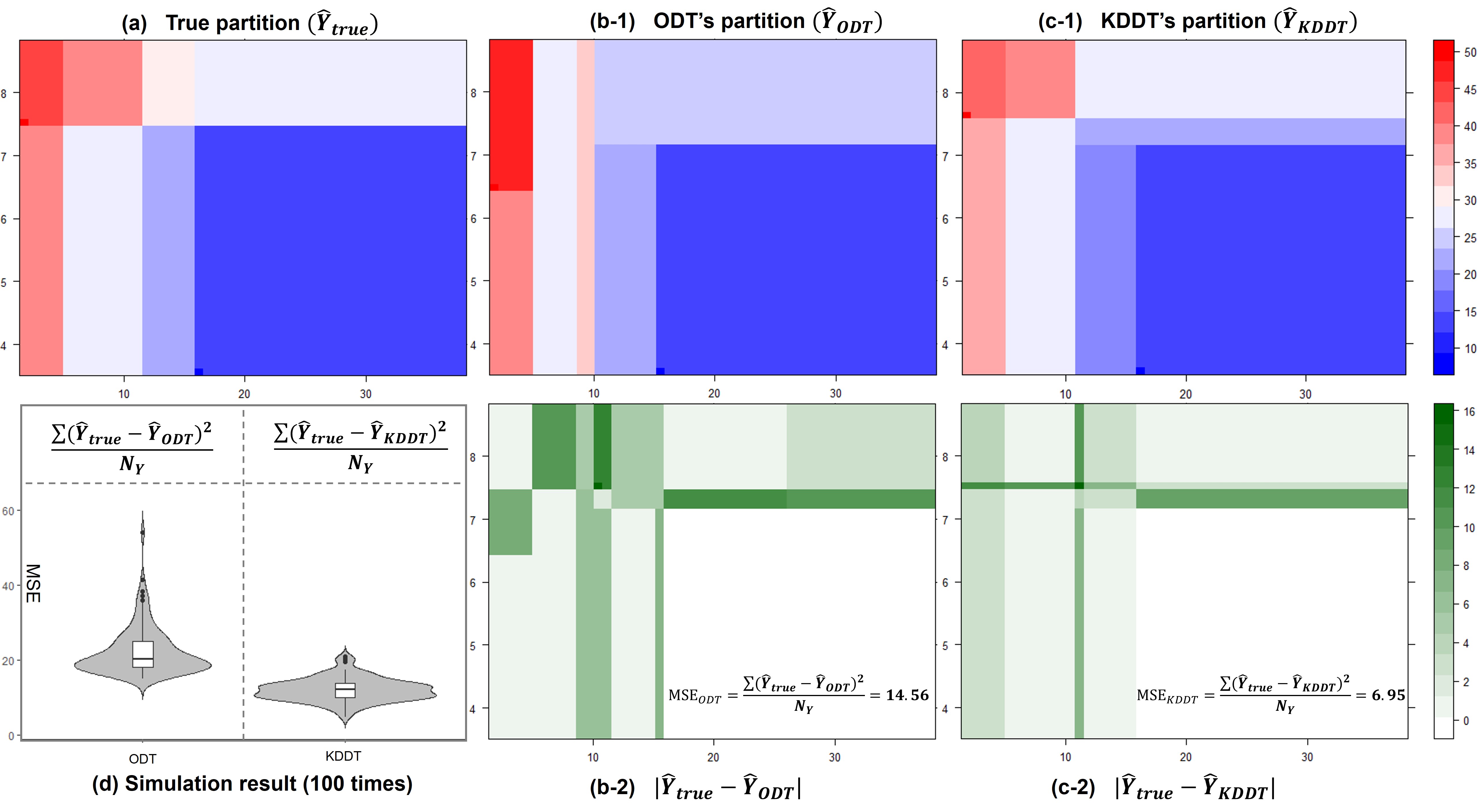}
\caption{\small The comparison of interpretations and simulation result. (a) The true partition is obtained from the ODT that is fitted based on the entire data of the true function. (b-1) The ODT is fitted based on the 50 samples. (b-2) The result of $|\hat{Y}_{true}-\hat{Y}_{ODT}|$. (c-1) The KDDT is built from RF. (c-2) The result of $|\hat{Y}_{true}-\hat{Y}_{KDDT}|$. (d) MSE comparison of ODT and KDDT with 100 times simulations.}
\label{fig:simInterpretComp}
\end{figure*}

For the function $y=f(x_1,x_2)$, effective interpretation is visually demonstrated through a suitable partition of the response values $y$ based on the covariates $x_1$ and $x_2$, as shown in panel (a) of Figure \ref{fig:sim2D}. This partition comprises nine splits generated by the ODT in panel (c), which is fitted using the entire dataset of 2601 data points. We refer to this partition as the true partition, representing the optimal interpretation. Although the random forest model provides an accurate estimation of the true function, it cannot generate a partition for interpretation. The ODT (fitted with the 50 samples) is interpretable, but its interpretation (partition) is not accurate. In contrast, the KDDT's interpretation closely approximates the optimal one (true partition), which is better than the ODT's. 
This claim relies on visual inspection which is a qualitative approach. Figure \ref{fig:simInterpretComp} presents a quantitative method for comparing the quality of interpretation between ODT and KDDT. Panel (a) displays the true partition (optimal interpretation). The partitions of ODT and KDDT are depicted in panels (b-1) and (c-1), respectively. Panels (b-2) and (c-2) illustrate the absolute errors of ODT and KDDT compared to the truth. Clearly, visual inspection still leads to the same conclusion that KDDT's interpretation (partition) is superior to ODT's. More importantly, we can quantify this difference using MSE. In this example, KDDT's MSE is 6.95, significantly smaller than ODT's MSE of 14.56. Furthermore, we repeat this comparison 100 times. The result in panel (d) demonstrates that, in general, KDDT outperforms ODT in terms of interpretation quality measured by MSE. Note that the medians of MSE are 20.32 and 12.14 for ODT and KDDT, respectively. The corresponding means of MSE are 22.39 and 12.17. KDDT results in a 40.3\% reduction in the median and a 45.6\% reduction in the mean compared to ODT. The maximum MSE of KDDT is 20.93, corresponding to 53 percentile of ODT's. The maximum MSE of ODT is 54.06, which is more than 2.5 times higher than the KDDT's.

\begin{figure*}[ht]
\centering
\includegraphics[width=\textwidth]{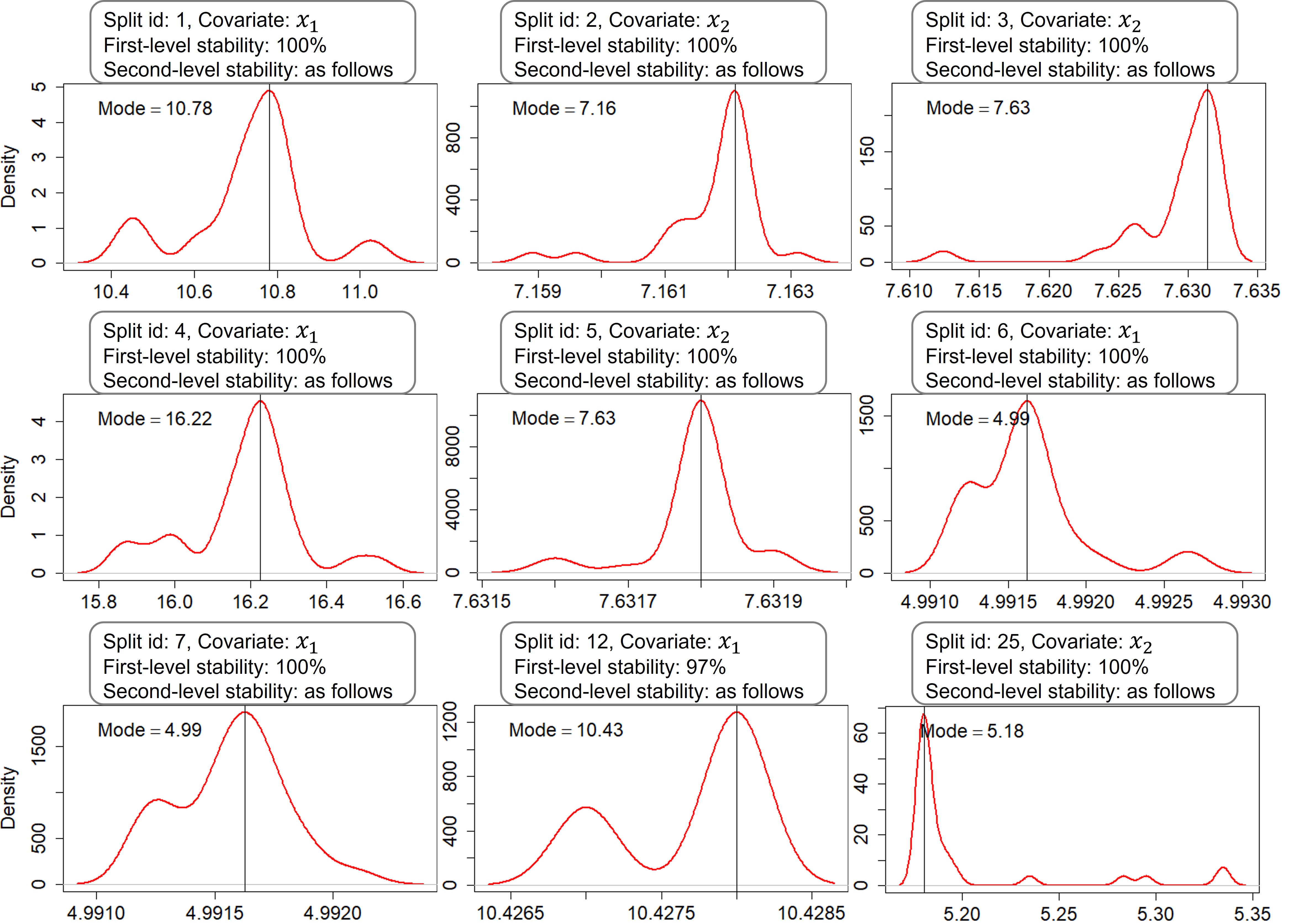}
\caption{\small The two-level stability of interpretable splits in panel (f) of Figure \ref{fig:sim2D}.}
\label{fig:simStability}
\end{figure*}

To examine the KDDT in Figure \ref{fig:sim2D} (c) in more details, it contains nine interpretable nodes (splits). The first-level and second-level stability can be found in Figure \ref{fig:simStability}. Except for split 12, which maintains a still impressive first-level stability of 97\%, all other splits exhibit a first-level stability of 100\%. Regarding second-level stability, each density function is tightly concentrated within a narrow interval and displays a sharp peak. Consequently, we can confidently assert that the interpretable splits within the KDDT are stable.

\section{Applications}
\label{sec:apply}
When should we use KDDT? Two fundamental conditions should be met.
\begin{itemize}
    \item \textbf{Demands for understanding or explanation:} We need to understand or explain the data, either to gain personal insight or to communicate findings to others.
    \item \textbf{Possess good prediction accuracy:} The black-box ML model, which KDDT aims to approximate, should outperform simple interpretable models, such as linear regression or ODT, in predicting the data. This suggests that the black-box model may have a better understanding of the data and the potential to offer a more accurate interpretation compared to the simple models. 
\end{itemize}

Considering these conditions, we discuss two real applications of KDDT in this section.

\subsection{Example for Model Interpretation}\label{sec:app-multi-interpretations}
In the application of model interpretation, we use the Boston Housing dataset, which comprises a total of 506 observations with 14 variables. The description of variables can be found in table \ref{tb:variables} in Appendix \ref{apdxB}. Our goal is to understand the effects of covariates on the price of houses in Boston (in 1970). To check the second condition, we select the linear regression model (LM) and ODT as simple interpretable models while choosing the random forest (RF) and SVM as two candidate black-box models. A five-fold cross-validation was conducted to compare their prediction accuracy. The MSE (mean square error) on testing data are LM: 23.2, ODT: 24.9, RF: 10.9, SVM: 13.4, and KDDT(RF): 14.9. More details of comparison can be found in Figure \ref{fig:realBHcompare} in Appendix \ref{apdxB}. From the results, the ML models outperform the simple interpretable models, and RF performs better than SVM. Hence, we can choose RF as the teacher model. The student model KDDT(RF) outperforms the simple interpretable models and exhibits similar performance to SVM. It indicates that KDDT(RF) may offer a more accurate interpretation than the simple interpretable models.

\begin{figure*}[t!]
\centering
\includegraphics[width=\textwidth]{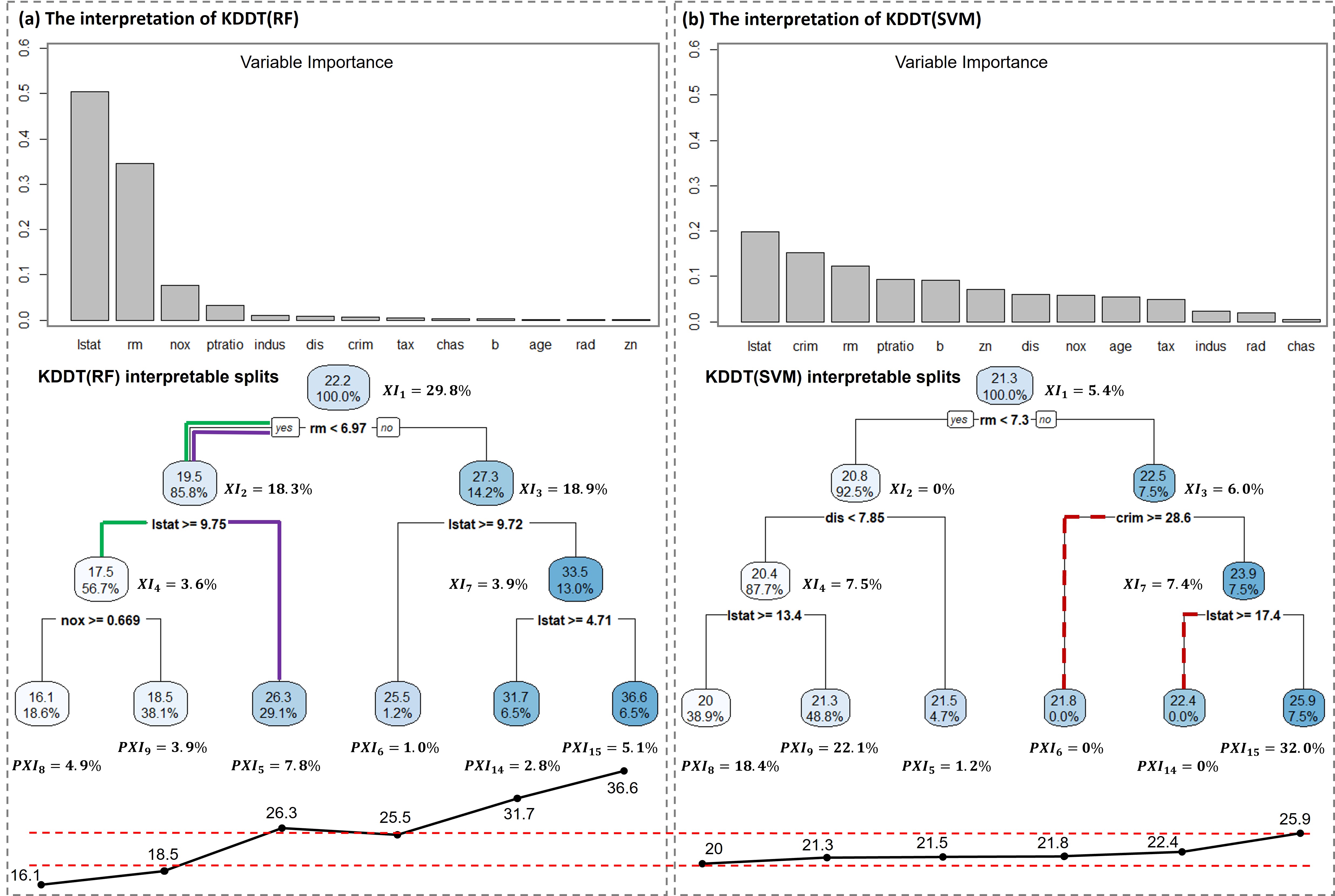}
\caption{\small Model interpretation through KDDT. (a) The interpretation of RF using KDDT(RF). (b) The interpretation of SVM using KDDT(SVM). Note that the left node (yes) and the right node (no) indicate whether the split condition is met or not, respectively. Note: the process of constructing KDDT(RF) in panel (a) can be found in the Figure \ref{fig:findHybridDDT} in Appendix \ref{apdxB}.}
\label{fig:realBHtree}
\end{figure*}

The panel (a) of Figure \ref{fig:realBHtree} illustrates the interpretations of KDDT(RF) for its teacher model RF. Since KDDT(RF) is essentially a decision tree, identifying the variables of importance is straightforward. The three most important variables are lstat, rm, and nox, related to social status, house size, and the natural environment, respectively. This is consistent with the corresponding results of the teacher model RF (see Figure \ref{fig:BHvarImp} in Appendix \ref{apdxB}), which is evidence that KDDT(RF) can provide accurate interpretation for its teacher model.
More detailed and specific interpretations can be obtained by examining the interpretable splits (nodes) featured in panel (a). For example, if a house has seven or more rooms and is situated in an affluent community where the percentage of the population with lower social status (lstat) is less than 4.71\%, it is likely to have a high value, averaging \$36,600. Additionally, for potential buyers, an intriguing insight emerges: they might acquire a larger house with seven or more rooms in a less affluent community with lstat $\geq 9.8\%$, priced around \$25,500, which is cheaper than a smaller house that could cost around \$26,300 in a community with lstat $\leq 9.7\%$. These specific insights are exemplified by nodes 5 and 6 in the tree. The stability of the interpretable splits shown in Figure \ref{fig:BHstability} in Appendix \ref{apdxB} ensures the credibility of interpretations.

\begin{figure*}[t!]\centering
\includegraphics[width=.9\textwidth]{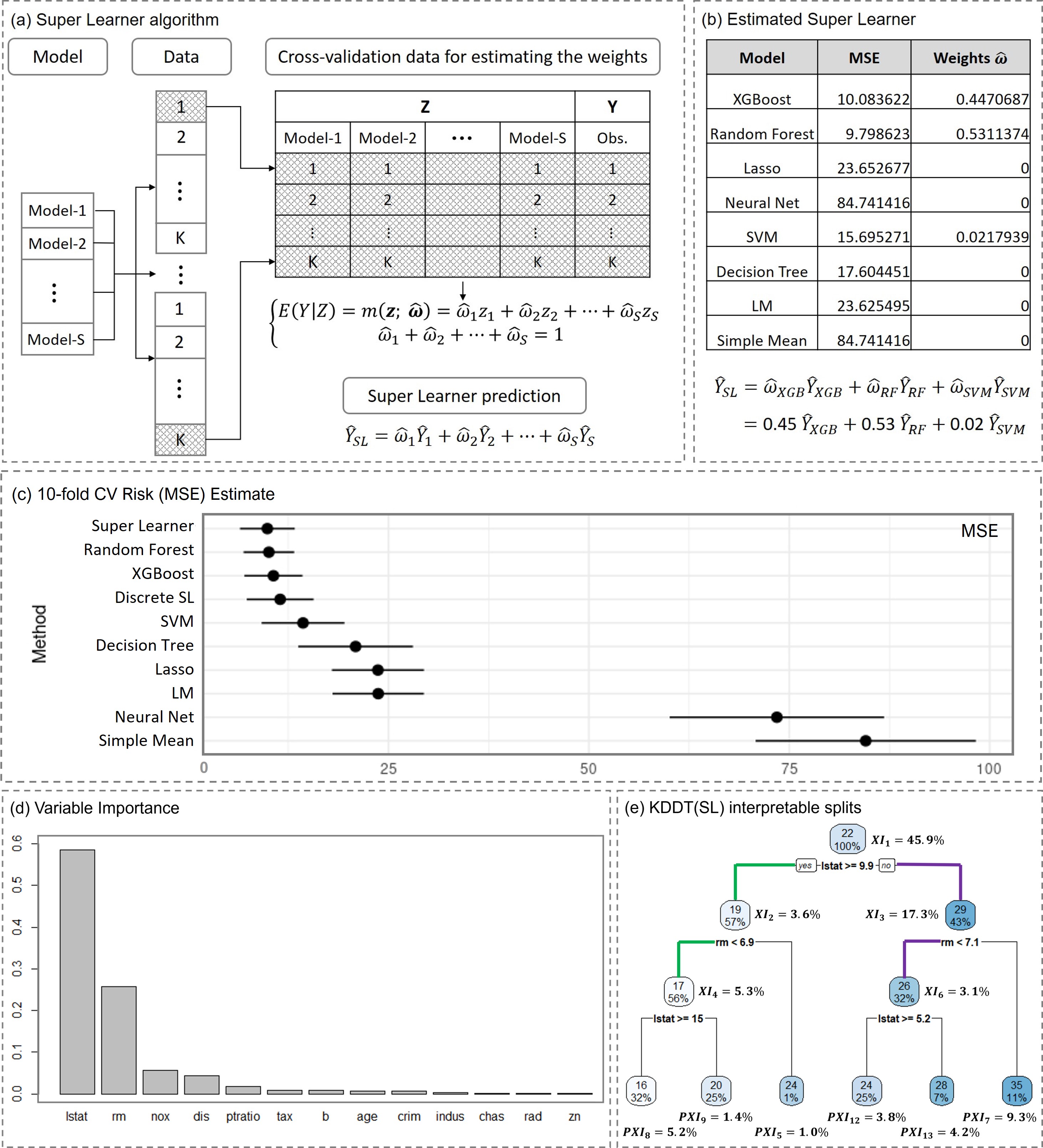}
\caption{\small The interpretation of super learner through KDDT. (a) The framework of super learner algorithm/model. The number $1,...,K$ refer to different cross-validation folds. The gray folds refers testing data. (b) The estimated weights and super learner. (c) The comparison of prediction accuracy between super learner and base models. (d) The variable importance of KDDT(SL). (e) The tree structure of KDDT(SL). }
\label{fig:SuperLearning}
\end{figure*}

In panel (a) of Figure \ref{fig:realBHtree}, the XI and PXI associated with the interpretable and predictive nodes provide the relative importance information for their interpretation. For instance, $XI_1=29.8\%$ for the split rm$<$6.97 indicates whether a house has seven or more rooms is crucial for assessing its value. Moreover, these indices could serve as stop criterion for identifying the interpretable nodes set. For example, we can identify the KDDT(RF) interpretable nodes by the criterion that the sum of PXI is less than 30\%. This criterion ensures that predictive nodes do not contain substantial information. Furthermore, we can interpret any prediction of KDDT by using the concept of the path explanation index in Definition \ref{def:XIP}. For example, if a prediction is made through the predictive node 9 (see panel (a)), its XI can be calculated as $XI_{1,9}=XI_1+XI_2+XI_4=51.7\%$. Then, with the $PXI_9=3.9\%$, we can obtain that $(\frac{XI_{1,9}}{XI_{1,9}+PXI_9},\frac{PXI_9}{XI_{1,9}+PXI_9})=(93\%,7\%)$. It indicates that the prediction can be interpreted with a degree of 93\% using the chain of decision rules $\{rm<6.97\xrightarrow{} lstat\geq 9.75\xrightarrow{}nox\geq 0.669\}$.

Last but not least, the percentage of observed data of each node also plays a pivotal role in comprehending the interpretation of KDDT. This percentage serves as crucial evidence of how strongly the interpretation of a particular node is supported by the observed data. Given that KDDT is not a direct interpretation of the observed data but rather of the teacher model, the support from the observed data is pivotal for the interpretation's practical significance. Even a node (split) with a high XI may lack practical relevance if the percentage of observed data associated with it (or its children) is exceedingly low. For instance, consider node 6 (split 3). Although it has $XI_{1,6}=XI_1+XI_3=48.7\%$ ($XI_3=18.9\%$), it (left child) comprises a mere 1.2\% of observed data.  This suggests that the interpretation of this node (split) might not carry much practical importance. In other words, the chance of purchasing a larger house at a lower price is not zero, but it is very low in practice. Consequently, it is imperative to take into account both the XI and the percentage of observed data when interpreting KDDT. As an example, we are confident in the interpretation of predictions made through node 9. Because this node not only has a high path XI of $XI_{1,9}=51.7\%$ that can be interpreted with a degree of 93\% but also enjoys strong practical support from a large number (38.1\%) of observed data.

As demonstrated in panel (b) of Figure \ref{fig:realBHtree}, KDDT can also provide an interpretation for SVM, which differs from the one for RF. In KDDT(SVM), the top three important variables are lstat, crim, and rm, related to social status, security, and house size, respectively. It indicates that, except for social status and house size, the SVM's explanation focuses on security, in contrast to RF emphasis on natural environment.
Regarding the interpretable splits, the sum of their XIs in KDDT(SVM) is 26.3\%, which is smaller than the 74.5\% in KDDT(RF). This suggests that the interpretable nodes set of KDDT(SVM) has less interpretability compared to its counterpart in KDDT(RF). Their comparison shown at the bottom of Figure 8 provides an intuitive illustration supporting this assertion, demonstrating that more variation in the data is explained by KDDT(RF) than by KDDT(SVM).
Another issue of KDDT(SVM) is that splits 3 and 7 have child nodes 6 and 14, respectively, which do not include any observed data. To address this, we can omit these two branches (red dashed lines) and focus solely on node 15. The path explanation index from node 1 to 15 can then be calculated as $XI_{1,15}=XI_1+XI_3+XI_7=18.8\%$. In sum, through KDDT, SVM can offer a different interpretation compared to RF. But, the interpretable splits in KDDT(SVM) do not perform as effectively as their counterparts in KDDT(RF).

KDDT can also be valuable in interpreting the model that is ensembled from other models. One typical example is the Super Learner introduced by \cite{vander2007}. As depicted in panel (a) of Figure \ref{fig:SuperLearning}, the Super Learner employs cross-validation to estimate the performance of multiple base models. Subsequently, it constructs an optimal weighted average of these models based on their testing performance. This approach has been proven to yield predictions that are asymptotically as good as or even better than any single model within the ensemble. In this example, we introduced eight base models and estimated their weights in the Super Learner, as shown in panel (b). Evaluated through a 10-fold cross-validation, the result presented in panel (c) demonstrates that the Super Learner outperforms all its base models in terms of prediction accuracy, which satisfies the second condition for applying KDDT.

Compared to the base models, the ensemble nature of the Super Learner renders it a more opaque black-box model, which makes the interpretation more challenging. KDDT can provide a solution. Panel (d) of Figure \ref{fig:SuperLearning} presents the variable importance of KDDT(SL), which remarkably resemble those of the RF model shown in panel (a) of Figure \ref{fig:realBHtree}. In panel (e) of Figure \ref{fig:SuperLearning}, interpretable splits (nodes) were selected based on the criterion that the sum of PXI is less than 30\%. The sum of XIs is 75.2\%, indicating that the interpretable nodes of KDDT(SL) offer substantial interpretability. An interesting observation emerges when comparing KDDT(RF) and KDDT(SL): the predictions and interpretations of nodes 4 and 5 in KDDT(RF) closely resemble those of nodes 4 and 6 in KDDT(SL). In Figure \ref{fig:realBHtree} (a) and Figure \ref{fig:SuperLearning} (e), these corresponding paths are highlighted in green and purple, respectively. Notably, all of the paths exhibit both high path XI and substantial percentages of observed data. This suggests a strong similarity in interpretation between RF and the Super Learner.

We have three KDDT interpretations associated with RF, SVM, and Super Learner. It's important to be aware that all of these interpretations are reasonable and valid. All roads lead to Rome. Choosing which one depends on the application requirements. For example, consider a real estate consultant whose client is interested in the natural environment of the house, KDDT(RF)'s explanation would be a good choice. If the client's main concern is the safety of the neighborhood, KDDT(SVM)'s interpretation may be a better choice. Furthermore, if significant splits or paths consistently appear in different KDDT interpretations, it serves as an indicator of their critical roles in the data. These interpretations have the potential to provide valuable insights or knowledge about the data or application. For example, as discussed in the comparison of KDDT(RF) and KDDT(SL), we can derive the valuable insight that 10\% lower status of the population and 7 rooms are two critical thresholds shaping people's evaluations of house prices in Boston. 

\subsection{Example for Subgroup Discovery}\label{sec:app-subgroup}

With the ability to uncover patterns in complex data and explore non-linear relationships, ML models have gained popularity in data-driven precision medicine, fueled by the rapid expansion in the availability of a wide variety of patient data. In precision medicine, identifying heterogeneity plays a central role, where subgroups of patients are defined based on baseline values of demographic, clinical, genomic, and other covariates, known as biomarkers. Understanding the effects of biomarkers in data analysis models is crucial for subgroup discovery. KDDT can bridge the gap between understanding the role of biomarkers and the lack of interpretability in black-box ML data analysis models. Particularly, as a tree-based approach, KDDT can incorporate information on higher-order interaction effects and be applied to define subgroups based on multiple biomarkers. Moreover, cutoff values do not need to be pre-specified for continuous/ordinal biomarkers. They are automatically estimated from the process of constructing KDDT.

\begin{figure*}[t!]
\centering
\includegraphics[width=.95\textwidth]{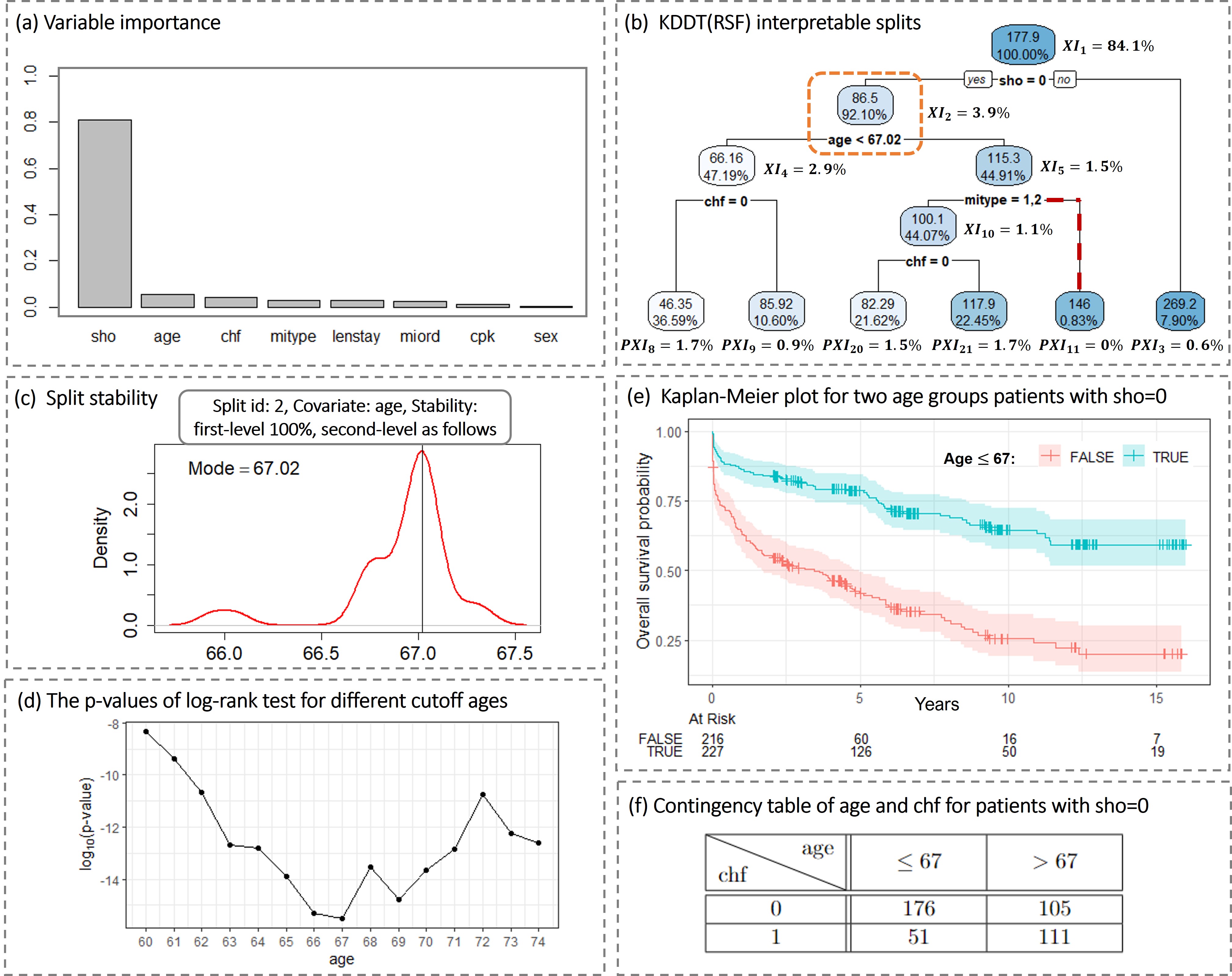}
\caption{\small Subgroup discovery and optimal cutoff identification. (a) Variable importance provides information for selecting split variables. (b) Interpretable splits for identifying optimal cutoff and subgroups. (c) The two-level stability of split 2. (d) Log10(p-value) of log-rank tests for validating the optimality of cutoff value. (e) The Kaplan-Meier plot for the identified subgroups. (f) Contingency table depicts the association between age and chf.}
\label{fig:realWhasTree}
\end{figure*}

In this example, we select the time-to-event dataset WHAS (Worcester Heart Attack Study), whose aim is to describe factors associated with trends in incidence and survival over time after admission for acute myocardial infarction. This dataset is available in the R package ``mlr3proba", and includes 481 observations and 14 variables. Four variables, id (Patient ID), year (Cohort year), yrgrp (Grouped cohort year), and dstat (Discharge status from the hospital: 1 = Dead, 0 = Alive), were excluded as they were not pertinent to the goal of study. The description of the remaining variables can be found in Table \ref{tb:whasVars} in Appendix \ref{apdxB}. We choose Cox Proportional Hazard (CoxPH) model as the interpretable model and Random Survival Forest (RSF) as the black-box teacher model. Similar to Section \ref{sec:app-multi-interpretations}, the comparison of prediction accuracy was conducted with a five-fold cross-validation. Instead of MSE, the C-index serves as the criterion, with higher C-index indicating higher accuracy. The result on testing data is CoxPH: 0.766, RSF: 0.797, and KDDT(RSF): 0.797. More details of the comparison can be found in Figure \ref{fig:realWhasCompare} in Appendix \ref{apdxB}. This result demonstrates the superiority of RSF in prediction and suggests that it is worth trying to take advantage of KDDT(RSF) in the application of subgroup discovery. 

The structure of KDDT(RSF) is depicted in panel (b) of Figure \ref{fig:realWhasTree}. As the first split, sho=0, $XI_1=84.1\%$ suggests a great practical significance for the identified subgroups. Actually, it is widely recognized that cardiogenic shock is positively associated with an increased risk of death. It is not a surprising discovery. The researcher's interest may lie more in the subgroups identified from the patients who didn't experience cardiogenic shock. The second split, age$<67.02$, reveals two subgroups. Although $XI_2=3.9\%$ is not high compared to $XI_1=84.1\%$, it is relatively high in the rest of interpretable nodes, $\frac{XI_2}{XI_2+XI_4+XI_5+XI_{10}}=\frac{3.9\%}{3.9\%+2.9\%+1.5\%+1.1\%}=41.5\%$. Moreover, the observed data in its child nodes are substantial and well-balanced, indicating strong support from the observed data. They are evidence that indicates the importance of the subgroups identified by the second split. The split stability of the optimal cutoff value is displayed in panel (c). The greedy search algorithm ensures its optimality which is substantiated by the p-values from log-rank tests across different values in panel (d). Consequently, there is no need to explore multiple cutoff values, thus alleviating the multiplicity issues. Panel (e) displays the Kaplan-Meier plot for the two subgroups, illustrating the varying risks associated with each subgroup. Finer subgroups and covariates interactions can be explored by considering deeper splits. Since node 11 just contains 4 observations (0.83\% of the data), we can remove it and its parent node 5 (see the red dashed line). This can be achieved by redistributing these 4 observations to nodes 20 and 21 based on their chf values. As a result, four subgroups with the number of patients can be identified in the table in panel (f). Analyzing this table reveals a clear interaction (dependency) between the risk of left heart failure (chf=1) and the age of patients. This relationship can be statistically confirmed through a $\chi^2$ test, which yields a p-value of 2.655e-10.

\section{Discussion}\label{sec:Dis}
KDDT offers a general method for interpreting black-box ML models, enabling the exploration of intricate data structures captured by these models for more precise and detailed interpretations. Essential attributes for good interpretable models include simplicity, stability, and predictivity. Stability is the central focus of this study. The primary challenge lies in constructing a stable KDDT while handling the randomness of the pseudo-data (knowledge) sampled in the knowledge distillation process. We propose a comprehensive theory for split stability and develop efficient algorithms for constructing stable KDDTs. To ensure simplicity, KDDT efficiently decouples the tasks of interpretation and prediction, maintaining a concise set of interpretable nodes for the purpose of interpretation. Regarding predictivity, KDDT, as a closed approximation of black-box ML models, retains strong predictive performance comparable to the original black-box models. In conclusion, KDDT is an excellent interpretable model with great potential for practical applications.

In our theory and algorithms, we employed the random sampling method to generate pseudo-data for constructing KDDT. This approach performed well in simulation and real data studies. Specifically, when the sample size is less than 60000, the time required to fit an interpretable node was under one minute. In general, for cases where the number of continuous covariates ($n_{con}$) is relatively small, typically less than 20, the sample size of 60000 is sufficient. However, when dealing with larger $n_{con}$, a larger sample size is necessary. In such cases, random sampling will be less efficient, and non-uniform random sampling strategies may be more attractive. Two promising strategies are MCMC sampling, which leverages information from the teacher model to enhance sampling efficiency, and PCA sampling, which uses dimension reduction to improve sampling efficiency. They are interesting directions for future study.

\begin{appendix}

\renewcommand\thefigure{\thesection.\arabic{figure}}
\renewcommand\thetable{\thesection.\arabic{table}}

\section{Theorems and Proofs}\label{apdxA}

Although the teacher model $f(\mathbf{x})$ may have $p$-dimensional covariates $\mathbf{x}=( x_1,...x_p)$, only one covariate is used at each split. Therefore, we need to marginalize over all other covariates to eliminate their influence. The result is a unary function $f(x)$ defined as follows.
    \begin{equation}
      \label{integral}
      f(x) = f_k(x_k) = \int ... \int f(\mathbf{x}) \,d\mathbf{x}_{-k}, \hspace{2mm} k = 1,...,p,
    \end{equation}
    where $\,d\mathbf{x}_{-k} = \prod_{i \neq k} \,dx_i$. If $x_j$ is categorical variables takes values in $\{1,...,C\}$, we set $\int f(\mathbf{x}_{-j},x_j) \,dx_j = \sum^C_{l=1} f(\mathbf{x}_{-j},x_j)I(x_j=l)$, where $\mathbf{x}_{-j}$ is the vector $\{x_i\}_{i \neq j}$, $I(\cdot)$ is an indicator function. We don't need to explicitly perform the integral for the univariate projection. It is implicitly handled in the greedy search algorithm by assessing the relationship between the univariate $x$ and the response $y$ at each split.

\begin{lemma}
\label{lemma}
Assume $x\in[a,b]$, where $a,b\in\mathbb{R}$, be a continuous variable in the teacher model $y = f(x)$, and $y$ can be a continuous or categorical variable. Let $z^l_c(x)=\int_{a}^{x} h^l_c(t) \,dt$, $z^r_c(x)=\int_{x}^{b} h^r_c(t) \,dt$, where $h^l_c(\cdot)$ and $h^r_c(\cdot)$ are integrable functions in $[a,b]$, $c\in\{1,...,C\}$ and $C\in \mathbb{N^+}$. The function $g(\cdot): \mathbb{R^C}\to \mathbb{R}$ is defined in Definition \ref{df-os}. Let $x_s$ be the unique optimal split in $(a,b)$ that is defined by \eqref{def-1}.

Consider $\{x_1,...,x_{n-1}\}$ as $n-1$ points drawn uniformly at random from the interval $(a,b)$, and arrange them in ascending order. Let $x_0=a$ and $x_n=b$, and include them in $\{x_1,...,x_{n-1}\}$ to form the set $\{x_0,x_1,...,x_{n-1}, x_n\}$. Utilizing the teacher model, we can generate pseudo-data as $\{(x_0,f(x_0)),(x_1,f(x_1)),...,(x_{n-1},f(x_{n-1})), (x_n,f(x_n))\}$. Subsequently, we can fit a stump to the pseudo-data by employing the greedy split search algorithm. The split criterion is defined as follows.

\begin{equation}\label{lemma-1}
\begin{split}
      x^n_s=  \underset{x_k,k \in \{1,...,n-1\}}{\operatorname{argmin}}[ & g(z^{l(n)}_1(x_k),...,z^{l(n)}_C(x_k)) + g(z^{r(n)}_1(x_{k+1}),...,z^{r(n)}_C(x_{k+1}))],
\end{split}
\end{equation}
where, $z^{l(n)}_c(x_k)=\sum_{i=1}^{k}h^l_c(x_i)* \Delta_i$, $z^{r(n)}_j(x_{k+1})=\sum_{j=k+1}^{n}h^r_c(x_j)* \Delta_j$
and $ \Delta_i = x_i-x_{i-1}, i=1,...,n$.

Let $k^n_s$ denote the optimal integer $k$ that minimized \eqref{lemma-1}, in other world, $x^n_s=x_{k^n_s}$. Then, the following holds:
$$x^n_s \overset{p}{\to} x_s, \hspace{3mm} as \hspace{2mm} n \to \infty .$$
The rate of convergence is $O(n^{-1})$.
\end{lemma}

\begin{proof}
There must exist a point $x_m$ such that, 
$$|x_s-x_m| = min\{|x_s-x_i|\}, \hspace{0.2cm} i=1,...,n-1.$$

Because $x_i, \hspace{0.2cm} i=1,...,n-1$ are uniformly distributed in $(a,b)$. For a constant $\epsilon, \hspace{0.2cm} 0<\epsilon<b-a$, by the theory of order statistics, it is easy to prove the following:
\begin{equation}
  \label{eq:oderStat}
P(|x_s-x_m|>\frac{\epsilon}{2})=(1-\frac{\epsilon}{b-a})^{n-1}.
\end{equation}
For any $\epsilon, \hspace{0.2cm} 0<\epsilon<b-a$,
\begin{equation}
  \label{lemma-2}
  \lim_{n \to \infty}P(|x_s-x_m|>\frac{\epsilon}{2})=\lim_{n \to \infty}(1-\frac{\epsilon}{b-a})^{n-1}=0.
\end{equation}
In other words, $x_m \overset{p}{\to} x_s$ as $n \to \infty$.

For any two consecutive points $x_{i-1}$ and $x_{i}$, $i=1,...,n$, it is easy to know
\begin{equation*}
    \begin{split}
        |x_i-x_{i-1}|= & min\{  |x_0-x_{i}|,\ldots,|x_{i-2}-x_{i}|,\\
         & |x_{i-1}-x_i|,|x_{i-1}-x_{i+1}|, \ldots,|x_{i-1}-x_{n}|\}.
    \end{split}
\end{equation*}
Thus, for a constant $\epsilon, \hspace{0.2cm} 0<\epsilon<b-a$, by the theory of order statistics, we have that
$$P(|x_i-x_{i-1}|>\frac{\epsilon}{2})=(1-\frac{\epsilon}{b-a})^{n-1}, \hspace{2mm} i=1,\ldots,n.$$
Let $\Delta_i=|x_i-x_{i-1}|$. Similar to \eqref{lemma-2}, we can prove that $\Delta_i \overset{p}{\to} 0$ as $n \to \infty$ for $i=1,...,n$.

Because $h(\cdot)$ is integrable in $[a,b]$ and $x_m \overset{p}{\to} x_s$, $\Delta_i \overset{p}{\to} 0$, as $n \to \infty$, we can get that
\begin{equation}\label{lemma-3}
\begin{split}
    & \lim_{n \to \infty}z^{l(n)}_c(x_m) \\ &=\lim_{n \to \infty}\sum_{i=1}^{m}h^l_c(x_i)* \Delta_i \overset{p}{\to} \int_{a}^{x_s} h^l_c(u) \,du \\
    &=z^l_c(x_s).
\end{split}
\end{equation}
Similarly, by $x_{m+1} \overset{p}{\to} x_s$ as $n \to \infty$ (because $\Delta_{m+1} \overset{p}{\to} 0$ as $n \to \infty$), we can prove that,
$$\lim_{n \to \infty}z^{r(n)}_c(x_{m+1})\overset{p}{\to}z^r_c(x_s).$$
Then, by the continuity of function $g$, we can obtain that
\begin{equation}
\label{lemma-4}
\begin{split}
& \lim_{n \to \infty}[g(z^{l(n)}_1(x_m),...,z^{l(n)}_C(x_m))  + g(z^{r(n)}_1(x_{m+1}),...,z^{r(n)}_C(x_{m+1}))] \\
     & \overset{p}{\to} g(z^l_1(x_s),...,z^l_C(x_s))+g(z^r_1(x_s),...,z^r_C(x_s)).
\end{split}
\end{equation}

Because of $\Delta_i=|x_i-x_{i-1}| \overset{p}{\to} 0$ as $n \to \infty$, for any two consecutive points, we have  $|x_{k^n_s}-x_{k^n_s+1}| \overset{p}{\to} 0$. Therefore, there must exist a point $x^*_s$, such that $x_{k^n_s} \overset{p}{\to} x^*_s$ and $x_{k^n_s+1} \overset{p}{\to} x^*_s$. Consequently, recall equation \eqref{lemma-1}, we have $x_{k^n_s}=x^n_s \overset{p}{\to} x^*_s$ as $n \to \infty$. Now, let us assume that the sequence $\{x^n_s\}$ does not converge solely to $x^*_s$. This would imply the existence of at least two distinct values, say $x^*_{s1}$ and $x^*_{s2}$ (with $x^*_{s1} \neq x^*_{s2}$), such that both $x^n_s \overset{p}{\to} x^*_{s1}$ and $x^n_s \overset{p}{\to} x^*_{s2}$ hold. This means that both $x^*_{s1}$ and $x^*_{s2}$ satisfied equation \eqref{lemma-1}, which is equivalent to the definition of an optimal split in equation \eqref{def-1}. This implies the existence of two optimal splits, contradicting the assumption of a unique optimal split.
So, we have 
$$x^n_s \overset{p}{\to} x^*_s, \hspace{3mm} as \hspace{2mm} n \to \infty .$$

Because $x^n_s=x_{k^n_s}$ and $x_{k^n_s+1} \overset{p}{\to} x^n_s$ as $n\to \infty$, we know
$$x_{k^n_s} \overset{p}{\to} x^*_s \hspace{3mm} , \hspace{3mm} x_{k^n_s+1} \overset{p}{\to} x^*_s.$$

Using the integrability of $h(\cdot)$ and the same proof procedures of \eqref{lemma-3} and \eqref{lemma-4}, we can obtain the following:

\begin{equation}
\label{lemma-6}
\begin{split}
& \lim_{n \to \infty}[g(z^{l(n)}_1(x_{k^n_s}),...,z^{l(n)}_C(x_{k^n_s}))  + g(z^{r(n)}_1(x_{k^n_s+1}),...,z^{r(n)}_C(x_{k^n_s+1}))] \\
     & \overset{p}{\to} g(z^l_1(x^*_s),...,z^l_C(x^*_s))+g(z^r_1(x^*_s),...,z^r_C(x^*_s)).
\end{split}
\end{equation}

According to the greedy search algorithm and split criterion in \eqref{lemma-1}, for all $n \in \mathbb{N}$, we know that
\begin{equation}
\label{lemma-7}
\begin{split}
& g(z^{l(n)}_1(x_m),...,z^{l(n)}_C(x_m))  + g(z^{r(n)}_1(x_{m+1}),...,z^{r(n)}_C(x_{m+1})) \\
     & \geq g(z^{l(n)}_1(x_{k^n_s}),...,z^{l(n)}_C(x_{k^n_s}))  + g(z^{r(n)}_1(x_{k^n_s+1}),...,z^{r(n)}_C(x_{k^n_s+1})).
\end{split}
\end{equation}
The equal sign holds if and only if $m=k^n_s$.

From \eqref{lemma-4}, \eqref{lemma-6}, and \eqref{lemma-7} we can get,
\begin{equation*}
\begin{split}
  g(z^l_1(x_s),...,z^l_C(x_s))+g(z^r_1(x_s),...,z^r_C(x_s)) 
 \geq g(z^l_1(x^*_s),...,z^l_C(x^*_s))+g(z^r_1(x^*_s),...,z^r_C(x^*_s)).
\end{split}
\end{equation*}

Since $x_s$ is the unique and optimal split that satisfies the split criterion \eqref{def-1}, the following must hold:
\begin{equation*}
\begin{split}
     g(z^l_1(x_s),...,z^l_C(x_s))+g(z^r_1(x_s),...,z^r_C(x_s)) \leq g(z^l_1(x^*_s),...,z^l_C(x^*_s))+g(z^r_1(x^*_s),...,z^r_C(x^*_s)).
\end{split}
\end{equation*}
So,
\begin{equation*}
\begin{split}
     g(z^l_1(x_s),...,z^l_C(x_s))+g(z^r_1(x_s),...,z^r_C(x_s))  = g(z^l_1(x^*_s),...,z^l_C(x^*_s))+g(z^r_1(x^*_s),...,z^r_C(x^*_s)),
\end{split}
\end{equation*}
and
\begin{equation*}
 x^n_s \overset{p}{\to} x^*_s=x_s, \hspace{3mm} as \hspace{2mm} n \to \infty. 
\end{equation*}

For the rate of convergence, let's recall \eqref{lemma-2} and change $\epsilon$ to $\frac{\epsilon'}{n}$, where $\epsilon'$ is a constant in $(0,b-a)$.
$$\lim_{n \to \infty}P(|x_s-x_m|>\frac{\epsilon'}{2n})=\lim_{n \to \infty}(1+\frac{-\epsilon'/(b-a)}{n})^{n}=e^{-\epsilon'/(b-a)}$$

Because $e^{-\epsilon'/(b-a)}$ is a constant, the convergence rate of $x_m$ is $O(n^{-1})$.

Since $|x_s-x^n_s|\geq |x_s-x_m|$ always holds $x^n_s$ converges to $x_s$ slower or equal to $x_m$.
However, by \eqref{lemma-7}, we know that 
$$g(z^{l(n)}_1(x_{k^n_s}),...,z^{l(n)}_C(x_{k^n_s})) + g(z^{r(n)}_1(x_{k^n_s+1}),...,z^{r(n)}_C(x_{k^n_s+1}))$$
converges to
$$g(z^l_1(x_s),...,z^l_C(x_s))+g(z^r_1(x_s),...,z^r_C(x_s))$$
faster or equal than 
$$g(z^{l(n)}_1(x_m),...,z^{l(n)}_C(x_m)) + g(z^{r(n)}_1(x_{m+1}),...,z^{r(n)}_C(x_{m+1}))$$
in all instances. This implies $x^n_s$ converges to $x_s$ faster or equal than $x_m$.

So, the convergence rate of $x^n_s$ is exactly the same as $x_m$, and it is at the level $O(n^{-1})$ too.
\end{proof}

\begin{cor}
Let $d = \frac{3(b-a)}{2(n-1)}$. The interval $[x^n_s-d,x^n_s+d]$ approximates to the 95\% confidence interval of the true optimal split $x_s$ for large $n$.
\end{cor}

\begin{proof}
According to the proof of Lemma 1, for a constant $d\in (0, \frac{b-a}{2}]$, $P(|x_s-x_m|>d)$ represents the probability that all random samples in $\{x_1,...,x_{n-1}\}$ fall outside the interval $[x_s-d,x_s+d]$. Similar to equation \eqref{eq:oderStat}, we have
$$P(|x_s-x_m|>d)=(1-\frac{2d}{b-a})^{n-1}.$$
Thus, $P(|x_s-x_m|\leq d)=1-(1-\frac{2d}{b-a})^{n-1}$ is the probability that the true optimal split $x_s$ lies within the interval $[x_m-d,x_m+d]$. In other words, $[x_m-d,x_m+d]$ is the confidence interval of the true optimal split $x_s$ at a significance level of $1-\alpha=1-(1-\frac{2d}{b-a})^{n-1}$.

Because neither $x_s$ nor $x_m$ are known. Let us recall the proof of Lemma \ref{lemma}, where we know that $x^n_s \overset{p}{\to} x_s$ and $x_m \overset{p}{\to} x_s$ as $n \to \infty$, and both $x_m$ and $x^n_s$ converge to $x_s$ at the same rate. Therefore, we have $(x^n_s \pm d) \overset{p}{\to} (x_s \pm d)$, $(x_m \pm d) \overset{p}{\to} (x_s \pm d)$, and $ |x_s-x^n_s| \overset{p}{\to} |x_s-x_m|$ as $n \to \infty$. Such that
\begin{equation*}
    \begin{split}
    & \lim_{n \to \infty}P(|x_s-x^n_s| \leq d)  \\
   = & \lim_{n \to \infty}P(|x_s-x_m| \leq d) \\
    = & \lim_{n \to \infty}[1-(1-\frac{2d}{b-a})^{n-1}] \\
       =&\lim_{n \to \infty}[1-(1+\frac{-3}{n-1})^{n-1}], \hspace{2mm} by \hspace{2mm} d = \frac{3(b-a)}{2(n-1)} \\
        =&1-e^{-3}\approx 0.950.
    \end{split}
\end{equation*}

So, the interval
\begin{equation}
  \label{theoreticalCI}
   [x^n_s-d,x^n_s+d]  \equiv [x^n_s-\frac{3(b-a)}{2(n-1)},x^n_s+\frac{3(b-a)}{2(n-1)}]
\end{equation}
is a good approximation of the 95\% confidence interval of the true optimal split $x_s$ for large $n$.
\end{proof}

\begin{figure*}[t!]\centering
\includegraphics[width=0.75\textwidth]{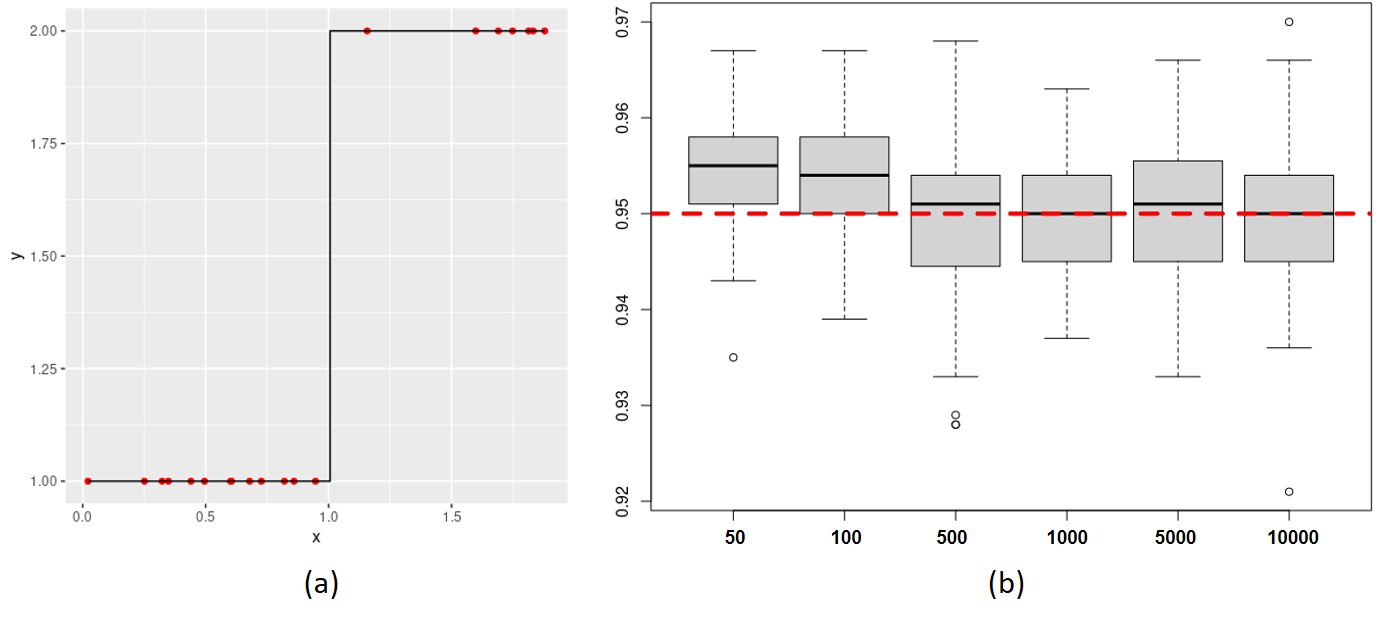}
\caption{(a) The true optimal split $x=1$. (b) The coverage rate of $95\%$ confidence intervals converges to the theoretical value (95\%).}
\label{fig:validateLemma1}
\end{figure*}

To empirically validate the theoretical results about the confidence interval \eqref{theoreticalCI}, we conducted an experiment using a step function, denoted as $f(x)$, with a single optimal split at $x=1$, as illustrated in panel (a) of Figure \ref{fig:validateLemma1}. The experiment proceeded as follows:
\begin{itemize}
    \item[(1)] At each sample size, we performed the following steps 1000 times:
    \begin{itemize}
        \item Calculated the theoretical 95\% confidence interval using \eqref{theoreticalCI}.
        \item Determined whether the true optimal split $x=1$ falls within this interval.
    \end{itemize}
    We then calculated the coverage rate, which is the proportion of times the true optimal split was covered by the confidence interval.
    \item[(2)] We repeated step (1) a total of 100 times to obtain the empirical distribution of the coverage rate.
\end{itemize}
The results of this experiment are presented in panel (b) of Figure \ref{fig:validateLemma1}. Notably, the empirical expectation of the coverage rate is approximately 95\% when the sample size is 500 or larger. This empirical finding aligns well with the theoretical 95\% confidence interval and supports the validity of confidence interval \eqref{theoreticalCI} and Lemma \ref{lemma} in a practical context.

\vspace{3mm} 
\begin{thm}[Continuous split convergence under the criterion of SSE]\label{th:scenario1}
Let $X$ be a continuous random variable that takes values in $[a,b]$, where $a,b\in\mathbb{R}$. The teacher model $f(x)$ is integrable in $[a,b]$. We assume the existence of an unknown unique optimal split $x_s$ in $(a,b)$, which is defined as follows:
\begin{equation}
  \label{th-1}
  x_s=\underset{x\in(a,b)}{\operatorname{argmin}}[\int_{a}^{x} (f(t)-\mu_l{(x)})^2 \,dt + \int_{x}^{b} (f(t)-\mu_r{(x)})^2 \,dt],
\end{equation}
where,
$$ \mu_l{(x)}= \frac{1}{x-a} \int_{a}^{x} f(u) \,du \hspace{0.5cm} , \hspace{0.5cm} \mu_r{(x)}=\frac{1}{b-x} \int_{x}^{b} f(u) \,du. $$

Consider $\{x_1,...,x_{n-1}\}$ as $n-1$ points drawn uniformly at random from the interval $(a,b)$, and arrange them in ascending order. Let $x_0=a$ and $x_n=b$, and include them in $\{x_1,...,x_{n-1}\}$ to form the set $\{x_0,x_1,...,x_{n-1}, x_n\}$. Utilizing the teacher model, we can generate pseudo-data as $\{(x_0,f(x_0)),(x_1,f(x_1)),...,(x_{n-1},f(x_{n-1})), (x_n,f(x_n))\}$. Subsequently, we can fit a stump to the pseudo-data by employing the greedy split search algorithm and splitting criterion SSE in \eqref{ssr}. Let $x^n_s$ denote the split of the stump.

Then, 1) $x^n_s$ converges to $x_s$ in probability as $n \to \infty$. 2) The values of two fitted nodes converge to $\mu_l(x_s)$ and $\mu_r(x_s)$ in probability respectively as $n \to \infty$. 3) The rate of convergence is $O(n^{-1})$.
\end{thm}

\begin{proof}
Construct 
$$g(z^l_1(x))=z^l_1(x), \hspace{2mm} g(z^r_1(x))=z^r_1(x),$$
$$z^l_1(x) = \int^x_a h^l(t) \,dt, \hspace{2mm} z^r_1(x) = \int^b_x h^r(t) \,dt,$$
$$h^l(t) = (f(t)-\mu_l(x))^2, \hspace{2mm} h^r(t) = (f(t)-\mu_r(x))^2,$$
$$z^{l(n)}_1(x_k)=\sum_{i=1}^{k}h^l(x_i)* \Delta_i \hspace{0.1cm}, \hspace{0.1cm} z^{r(n)}_1(x_{k+1})=\sum_{j=k+1}^{n}h^r(x_j)* \Delta_j \hspace{0.1cm},$$
where $\Delta_i = x_i-x_{i-1}, i=1,...,n$.

Under this construction, the optimal split $x_s$ defined in \eqref{th-1} follows \eqref{def-1} in Definition \ref{df-os}. 
Obviously, $h^l(\cdot)$ and $h^r(\cdot)$ are integrable in $[a,b]$, because $f(\cdot)$ is integrable. So, by applying Lemma \ref{lemma}, $x^n_s$ converges to $x_s$ in probability and the rate of convergence is $O(n^{-1})$.

Since $f(\cdot)$ is integrable in $[a,b]$ and $x_{k^n_s} = x^n_s$, $x_{k^n_s+1} \overset{p}{\to} x^n_s$, $x^n_s \overset{p}{\to} x_s$, as $n \to \infty$, we can prove that
\begin{equation*}
    \begin{split}
       & \lim_{n \to \infty}\frac{1}{x_{k^n_s}-a} \sum^{k^n_s}_{i=1}f(x_i) \Delta_i \overset{p}{\to} \frac{1}{x_s-a} \int_{a}^{x_s} f(u) \,du \\ &=\mu_l(x_s),
    \end{split}
\end{equation*}

\begin{equation*}
    \begin{split}
        & \lim_{n \to \infty}\frac{1}{b-x_{k^n_s+1}} \sum^{n}_{j=k^n_s+1}f(x_j) \Delta_j \overset{p}{\to} \frac{1}{b-x_s} \int_{x_s}^{b} f(u) \,du \\ &=\mu_r(x_s),
    \end{split}
\end{equation*}
and the rate of convergence is $O(n^{-1})$.
\end{proof}

\vspace{3mm} 
\begin{thm}[Continuous split convergence under the Tsallis entropy criterion]\label{th:scenario2}
Let $X$ be a continuous random variable that takes values in $[a,b]$, where $a,b\in\mathbb{R}$. $y = f(x)$ is the teacher model. $Y=f(X)$ is a discrete random variable taking values $y \in \{y_1,...,y_C\}$, where $C \in \mathbb{N^+}$. Let $S_i= \{x|f(x)=y_i, x\in [a,b]\}$, $i=1,...,C$. The probability mass function of $Y$ in $[a,b]$ is that:
$$p(y_i) =  \int_{S_i} \frac{1}{b-a} \,dx, \hspace{3mm} i=1,...,C.$$
And, the probability mass function of $Y$ in $[a,x]$ is that:
$$p_{[a,x]}(y_i)=  \int_{S_i \cap [a,x]} \frac{1}{x-a} \,dt.$$
Then, a Tsallis entropy can be calculated in $[a,x]$,
$$S_q([a,x])=\frac{1}{1-q}(\sum_{i=1}^{C}p_{[a,x]}(y_i)^q-1), \hspace{3mm} q\in \mathbb{R}.$$
Assume the existence of an unknown unique optimal split $x_s$ in $(a,b)$, which is defined as follows:
\begin{equation}
  \label{th-2}
   x_s=\underset{x\in(a,b)}{\operatorname{argmin}} [S_q([a,x]) +S_q([x,b])].
\end{equation}

Consider $\{x_1,...,x_{n-1}\}$ as $n-1$ points drawn uniformly at random from the interval $(a,b)$, and arrange them in ascending order. Let $x_0=a$ and $x_n=b$, and include them in $\{x_1,...,x_{n-1}\}$ to form the set $\{x_0,x_1,...,x_{n-1}, x_n\}$. Utilizing the teacher model, we can generate pseudo-data as $\{(x_0,f(x_0)),(x_1,f(x_1)),...,(x_{n-1},f(x_{n-1})), (x_n,f(x_n))\}$. Subsequently, we can fit a stump to the pseudo-data by employing the greedy split search algorithm and Tsallis entropy splitting criterion in \eqref{entropy} and \eqref{tsallis}. $x^n_s$ denotes the split of the stump.

Then, 1) $x^n_s$ converges to $x_s$ in probability as $n \to \infty$. 2) The rate of convergence is $O(n^{-1})$.
\end{thm}

\begin{proof}
Construct 
$$g(z^l_1(x),...,z^l_C(x))=\frac{1}{1-q}(\sum_{c=1}^{C}z^l_c(x)^q-1),$$
$$g(z^r_1(x),...,z^r_C(x))=\frac{1}{1-q}(\sum_{c=1}^{C}z^r_c(x)^q-1),$$
\begin{equation*}
    \begin{split}
        z^l_c(x) &= \int^x_a h^l_c(t) \,dt =\int^x_a \frac{1}{x-a} * I_{y_c}(f(t)) \,dt =\int_{S_i \cap [a,x]} \frac{1}{x-a} \,dt=p_{[a,x]}(y_c),
    \end{split}
\end{equation*}

\begin{equation*}
    \begin{split}
        z^r_c(x) &= \int^b_x h^r_c(t) \,dt =\int^b_x \frac{1}{b-x} * I_{y_c}(f(t)) \,dt =\int_{S_i \cap [x,b]} \frac{1}{b-x} \,dt=p_{[x,b]}(y_c),
    \end{split}
\end{equation*}
$$h^l_c(t) = \frac{1}{x-a} * I_{y_c}(f(t)), \hspace{2mm} h^r_c(t) = \frac{1}{b-x} * I_{y_c}(f(t)),$$
$$z^{l(n)}_c(x_k)=\sum_{i=1}^{k}h^l_c(x_i)* \Delta_i=\frac{1}{x-a}\sum_{i=1}^{k} \Delta_i * I_{y_c}(f(x_i)),$$
$$z^{r(n)}_c(x_{k+1})=\sum_{j=k+1}^{n}h^r_c(x_j)* \Delta_j = \frac{1}{b-x}\sum_{j=k+1}^{n}\Delta_j  * I_{y_c}(f(x_j)),$$
where $c=1,...,C$, $ \Delta_i = x_i-x_{i-1}, i=1,...,n$ and $I_{y_c}(f(x))$ is an indicator function that is equal to 1 at $f(x)=y_c$ and 0 elsewhere.

Under this construction, the optimal split $x_s$ defined in \eqref{th-2} follows \eqref{def-1} in Definition \ref{df-os}. 
Obviously, $h^l_c(\cdot)$ and $h^r_c(\cdot)$ are integrable in $[a,b]$. So, by applying Lemma \ref{lemma}, $x^n_s$ converges to $x_s$ in probability and the rate of convergence is $O(n^{-1})$.
\end{proof}

\vspace{3mm} 

\begin{thm}[Categorical split convergence under MSE criterion]\label{th:scenario3}
$X$ is a discrete random variable taking values $x \in \{1,...,C_x\}$, where $C_x \in \mathbb{N^+}$. $Y$ is a continuous random variable taking values $y \in [c,d]$, where $c,d \in \mathbb{R}$. $Y$ has a finite mean $\mu$. The conditional distribution of $Y|X=k$ is defined through the teacher model $y=f(k)$. The expectation of $Y|X=k$ is given by:
\begin{equation}\label{properties}
\begin{split}
    & E(Y|X=k)=\mu_k, \hspace{1mm}  \mu=\frac{1}{C_x}\sum^{C_x}_{k=1}\mu_k , \hspace{1mm} k=1,...,C_x .
\end{split}
\end{equation}

Let us randomly sample $n$ instances of $X$, denoted as $\{x_1,...,x_n\}$, from $\{1,...,C_x\}$. Corresponding samples of $Y$, denoted as $\{y_1,...,y_n\}$, are generated through the conditional distribution of $Y|X=k$. The uniform sampling assumption indicates $\lim_{n \to \infty}\frac{n_k}{n}=\frac{1}{C_x}$,
where $n_k=\sum^n_{i=1} I(x_i=k)$, $k=1,...,C_x$.

Assume the existence of an unknown unique optimal split $x_s$ in $\{1,...,C_x\}$, which is defined as follows:
\begin{equation}\label{th-3}
\begin{split}
       x_s = \underset{k\in\{1,...,C_x\}}{\operatorname{argmin}} \lim_{n \to \infty} \frac{1}{n}[ & \sum^{n_k}_{i=1}(y_{ki}-\mu_k)^2 +\sum_{l\neq k}(\sum^{n_l}_{j=1}(y_{lj}-\mu_{-k})^2)],
\end{split}
\end{equation}
where $\mu_{-k}=E(Y|X\neq k)$.

A stump can be fitted on the pseudo-data $\{(x_1,y_1),...,(x_n,y_n)\}$ by using the greedy search algorithm and splitting MSE criterion in \eqref{ssr}. Let $x^n_s$ denote the split of the stump.

Then, 1) $x^n_s$ converges to $x_s$ in probability as $n \to \infty$. 2) The rate of convergence is $O(n^{-1})$.

\end{thm}

\begin{proof}
Let's construct
\begin{equation*}
    \begin{split}
        & \lim_{n \to \infty} \frac{1}{n}\sum^{n_k}_{i=1}(y_{ki}-\mu_k)^2 = z^l(k), \hspace{3mm} \lim_{n \to \infty} \sum_{l\neq k}(\sum^{n_l}_{j=1}(y_{lj}-\mu_{-k})^2)=z^r(k), \hspace{3mm} g(z)=z.
    \end{split}
\end{equation*}
Obviously, \eqref{th-3} follows \eqref{def-1}, so $x_s$ follows Definition \ref{df-os}. 

By the splitting criterion \eqref{ssr}, the optimal split of the stump can be found that

\begin{equation*}
   x^n_s = \underset{k\in\{1,...,C_x\}}{\operatorname{argmin}} \frac{1}{n}[\sum^{n_k}_{i=1}(y_{ki}-\Bar{y}_k)^2+\sum_{l\neq k}(\sum^{n_l}_{j=1}(y_{lj}-\Bar{y}_{-k})^2)],
\end{equation*}
where $$\Bar{y}_k=\frac{1}{n_k}\sum^{n_k}_{i=1}y_{ki} \hspace{3mm} and \hspace{3mm} \Bar{y}_{-k}=\frac{1}{\sum_{l\neq k}n_l}\sum_{l\neq k}\sum^{n_l}_{j=1}y_{lj}.$$

Since we know that $n_k=\frac{1}{C_x}n, \hspace{3mm} k=1,...,C_x$. By the weak law of large numbers, we can prove that
$$\Bar{y}_k \overset{p}{\to} \mu_k,\hspace{3mm} \Bar{y}_{-k}\overset{p}{\to} \mu_{-k} \hspace{3mm} as \hspace{3mm} n \to \infty.$$

So, with probability one,
\begin{equation*}
\begin{split}
 \lim_{n \to \infty} x^n_s & = \underset{k\in\{1,...,C_x\}}{\operatorname{argmin}} \lim_{n \to \infty} \frac{1}{n}[ \sum^{n_k}_{i=1}(y_{ki}-\Bar{y}_k)^2  +\sum_{l\neq k}(\sum^{n_l}_{j=1}(y_{lj}-\Bar{y}_{-k})^2)] \\ 
 & \overset{p}{\to} \underset{k\in\{1,...,C_x\}}{\operatorname{argmin}} \lim_{n \to \infty} \frac{1}{n}[ \sum^{n_k}_{i=1}(y_{ki}-\mu_k)^2  +\sum_{l\neq k}(\sum^{n_l}_{j=1}(y_{lj}-\mu_{-k})^2)] = x_s.
\end{split}
\end{equation*}

$x^n_s$ converges to $x_s$ in probability and the rate of convergence is $O(n^{-1})$.
\end{proof}

\vspace{3mm} 

\begin{thm}[Categorical split convergence under Tsallis entropy criterion]\label{th:scenario4}
$X$ is a discrete random variable taking values $x \in \{1,...,C_x\}$, where $C_x \in \mathbb{N^+}$. $Y$ is a discrete random variable taking values $y \in \{1,...,C_y\}$, where $C_y \in \mathbb{N^+}$. A joint distribution $(X,Y)$ can be defined through the teacher model $y=f(x)$. Its probability mass function can be denoted as $p(x=i,y=j)=p_{ij}$, where $i=1,...,C_x,\hspace{3mm} j=1,...,C_y$.

Let us randomly sample $n$ instances of $X$, denoted as $\{x_1,...,x_n\}$, from $\{1,...,C_x\}$. Corresponding samples of $Y$, denoted as $\{y_1,...,y_n\}$, are generated through the joint distribution $(X,Y)$. The uniform sampling assumption indicates $\lim_{n \to \infty}\frac{n_k}{n}=\frac{1}{C_x}$, where $n_k=\sum^n_{i=1}I(x_i=k)$, $k=1,...,C_x$.

Assume the existence of an unknown unique optimal split $x_s$ in $\{1,...,C_x\}$, which is defined as follows:
\begin{equation}
  \label{th-4}
   x_s=\underset{k\in\{1,...,C_x\}}{\operatorname{argmin}} [S_q(k) +S_q(-k)],
\end{equation}
where, $S_q(\cdot)$ is the Tsallis entropy,
$$S_q(k)=\frac{1}{1-q}(\sum_{j=1}^{C_y}(p_{kj})^q-1), $$
$$S_q(-k)=\frac{1}{1-q}(\sum_{j=1}^{C_y}(\sum_{i\neq k} p_{ij})^q-1), \hspace{2mm} k,i\in \{1,...,C_x\},\hspace{2mm} q\in \mathbb{R}.$$

A stump can be fitted with the pseudo-data $\{(x_1,y_1),...,(x_n,y_n)\}$ by using the greedy search algorithm and the Tsallis entropy splitting criterion in \eqref{entropy} and \eqref{tsallis}. Let $x^n_s$ denote the split of the stump.

Then, 1) $x^n_s$ converges to $x_s$ in probability as $n \to \infty$. 2) The rate of convergence is $O(n^{-1})$.
\end{thm}

\begin{proof}

Let us construct
$$p_{kj} = z^l(k), \hspace{3mm} \sum_{i\neq k} p_{ij}=z^r(k) \hspace{3mm} and \hspace{3mm} g(z(k))=S_q(k).$$
Obviously, \eqref{th-4} follows \eqref{def-1}, so $x_s$ follows Definition \ref{df-os}. 

By the splitting criterion \eqref{entropy}, the optimal split of the stump can be found that
\begin{equation*}
   x^n_s=\underset{k\in\{1,...,C_x\}}{\operatorname{argmin}} [S^n_q(k) +S^n_q(-k)],
\end{equation*}
where
$$S^n_q(k)=\frac{1}{1-q}(\sum_{j=1}^{C_y}(\frac{1}{n_k}\sum^{n_k}_{l=1} I(y_l=j))^q-1), $$
\begin{equation*}
    \begin{split}
        & S^n_q(-k) = \frac{1}{1-q}(\sum_{j=1}^{C_y}(\sum_{i\neq k} (\frac{1}{n_i}\sum^{n_i}_{m=1} I(y_m=j)))^q-1), \hspace{3mm} i\in \{1,...,C_x\},\hspace{3mm} q\in \mathbb{R}.
    \end{split}
\end{equation*}

Since we know that $n_k=\frac{1}{C_x}n, \hspace{3mm} k=1,...,C_x$. By Borel's law of large numbers, with probability one,
$$ \lim_{n \to \infty} \frac{1}{n_k}\sum^{n_k}_{m=1} I(y_m=j)=p_{kj},\hspace{2mm} k=1,...,C_x ,\hspace{2mm} j=1,...,C_y.$$

So, with probability one,
\begin{equation*}
\begin{split}
 \lim_{n \to \infty} x^n_s & = \underset{k\in\{1,...,C_x\}}{\operatorname{argmin}} \lim_{n \to \infty} [S^n_q(k) +S^n_q(-k)] = \underset{k\in\{1,...,C_x\}}{\operatorname{argmin}} [S_q(k) +S_q(-k)] = x_s.
\end{split}
\end{equation*}

$x^n_s$ converges to $x_s$ in probability and the rate of convergence is $O(n^{-1})$.
\end{proof}

\onecolumn
\section{Supplementary materials}\label{apdxB}

\begin{figure*}[ht]
\centering
\includegraphics[width=.8\textwidth]{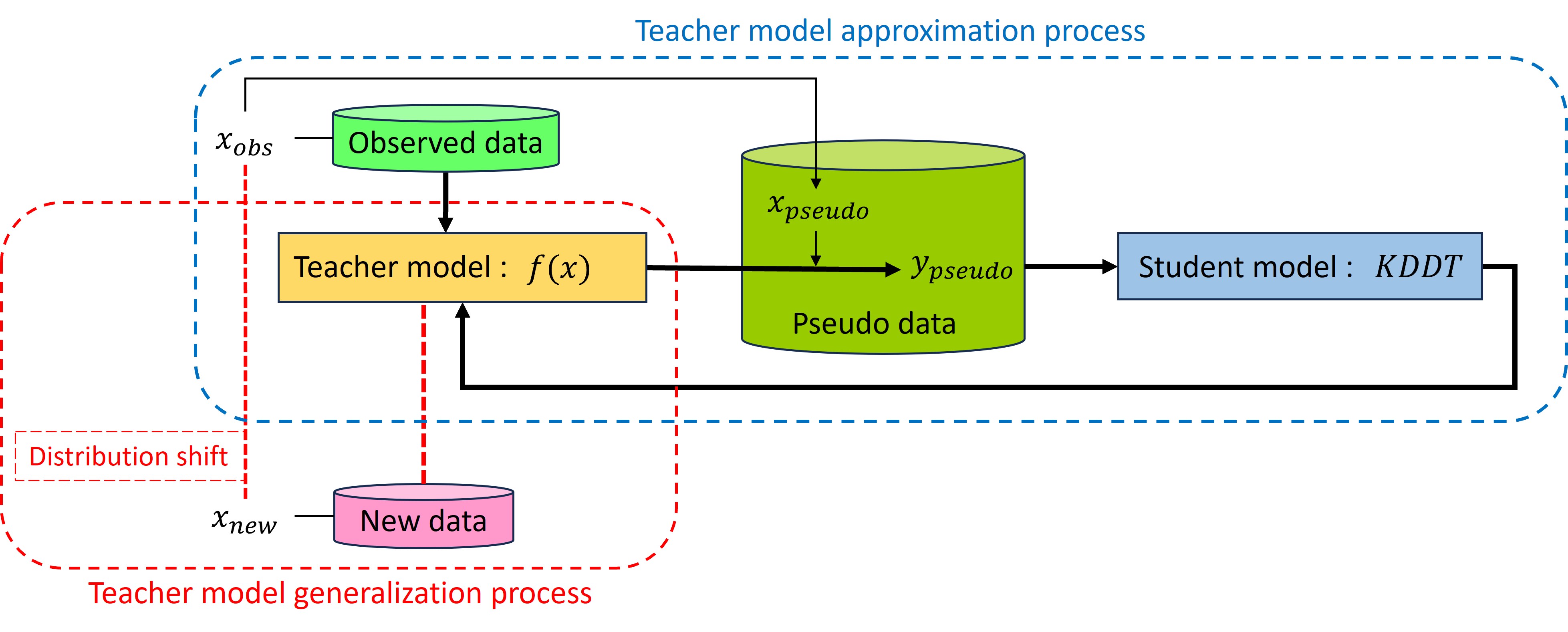}
\caption{\small The teacher model approximation and generalization process. The model generalization process may encounter the challenge of distribution shift, whereas the approximation process does not.}
\label{fig:approximation}
\end{figure*}

\begin{table*}[ht]
\renewcommand{\arraystretch}{1}
\caption{\label{tb:variables}Variables and short descriptions}
\small
\centering
\fbox{%
\begin{tabularx}{.7\textwidth}{|l|X|}
  \hline
  Variable & \multicolumn{1}{c|}{Short descriptions}  \\
  \hline \hline
  medv & median value of owner-occupied homes in USD 1000's. \\
  \hline
  crim & per capita crime rate by town. \\
  \hline
  zn & proportion of residential land zoned for lots over 25,000 sq.ft. \\
  \hline
  indus & proportion of non-retail business acres per town. \\
  \hline
  chas & Charles River dummy variable (= 1 if tract bounds river; 0 otherwise). \\
  \hline
  nox & nitric oxides concentration (parts per 10 million). \\
  \hline
  rm & average number of rooms per dwelling. \\
  \hline
  age & proportion of owner-occupied units built prior to 1940. \\
  \hline
  dis & weighted distances to five Boston employment centers. \\
  \hline
  rad & index of accessibility to radial highways. \\
  \hline
  tax & full-value property-tax rate per USD 10,000. \\
  \hline
  ptratio & pupil-teacher ratio by town. \\
  \hline
  b & $1000(B - 0.63)^2$ where B is the proportion of blacks by town. \\
  \hline
  lstat & percentage of lower status of the population. \\
  \hline
\end{tabularx}}
\end{table*}

\begin{figure*}[ht]
\centering
\includegraphics[width=.85\textwidth]{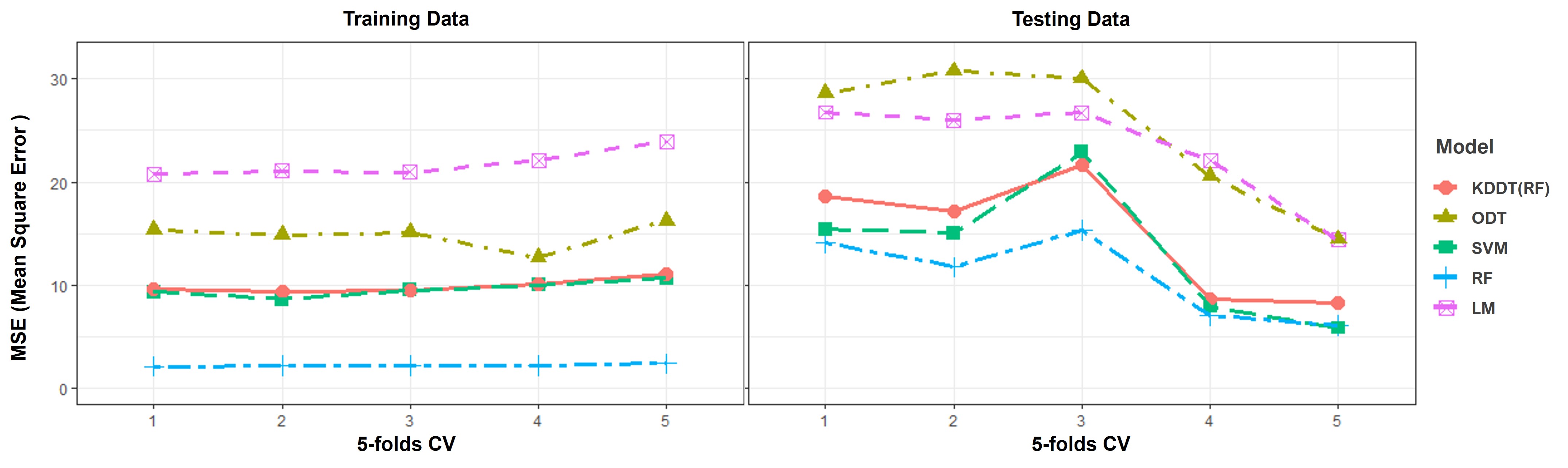}
\caption{\small Comparison of prediction accuracy among ODT, LM, SVM, RF, and KDDT(RF) on the training dataset (left) and testing dataset (right). Note that RF is the teacher model and KDDT(RF) is the student model. We included SVM to demonstrate that any black-box ML model can serve as the teacher model, and we opted for the one with higher prediction accuracy.}
\label{fig:realBHcompare}
\end{figure*}

\begin{figure*}[ht]
\centering
\includegraphics[width=.9\textwidth]{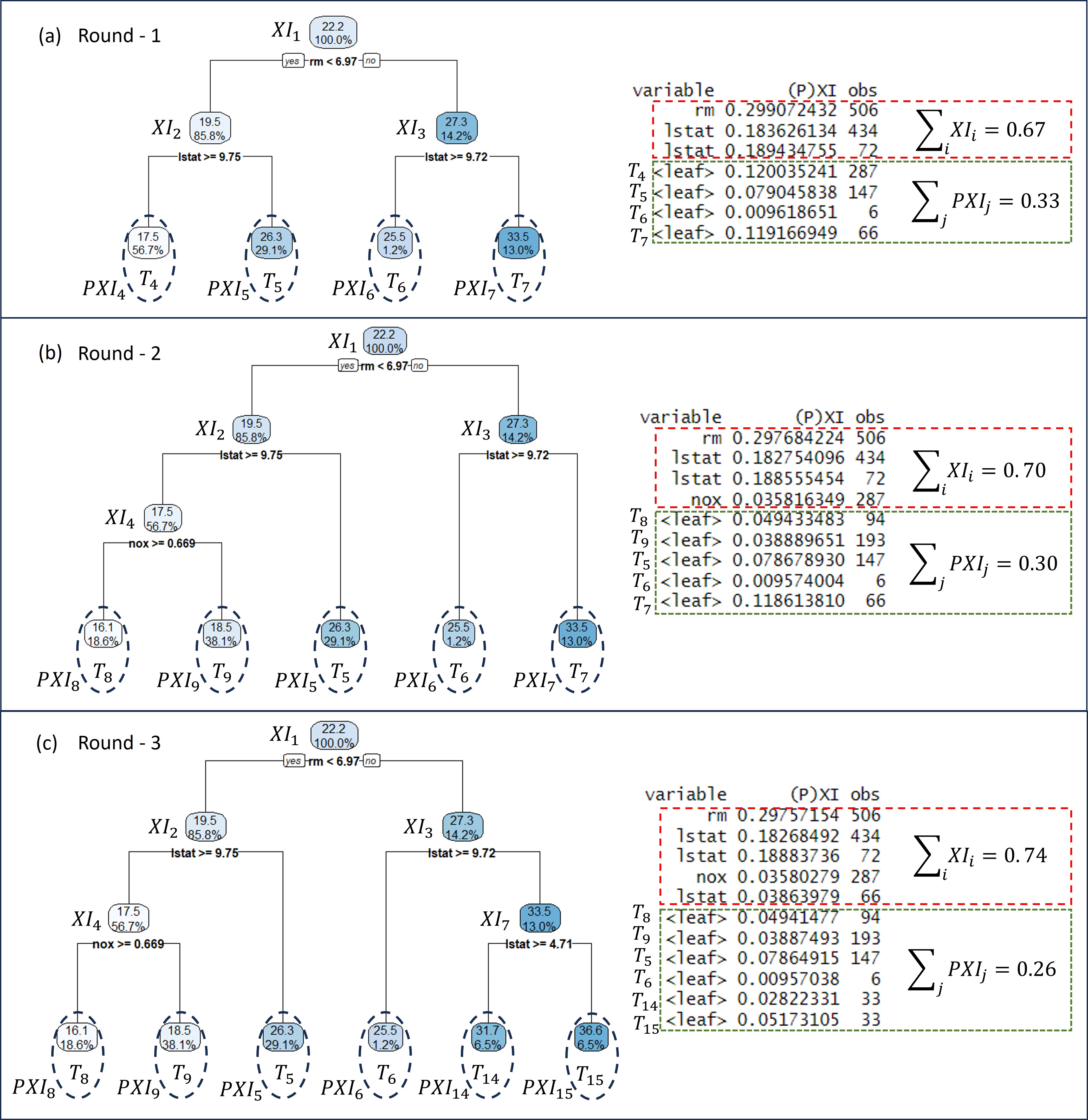}
\caption{\small An example to find the desired hybrid KDDT under the criterion of $\sum_i XI_i>70\%$ (or $\sum_j PXI_j<30\%$). The final hybrid KDDT in panel (c) is same with the one in the panel (a) of Figure \ref{fig:realBHtree}.}
\label{fig:findHybridDDT}
\end{figure*}

\FloatBarrier
\begin{figure*}[ht]
\centering
\includegraphics[width=0.7\textwidth]{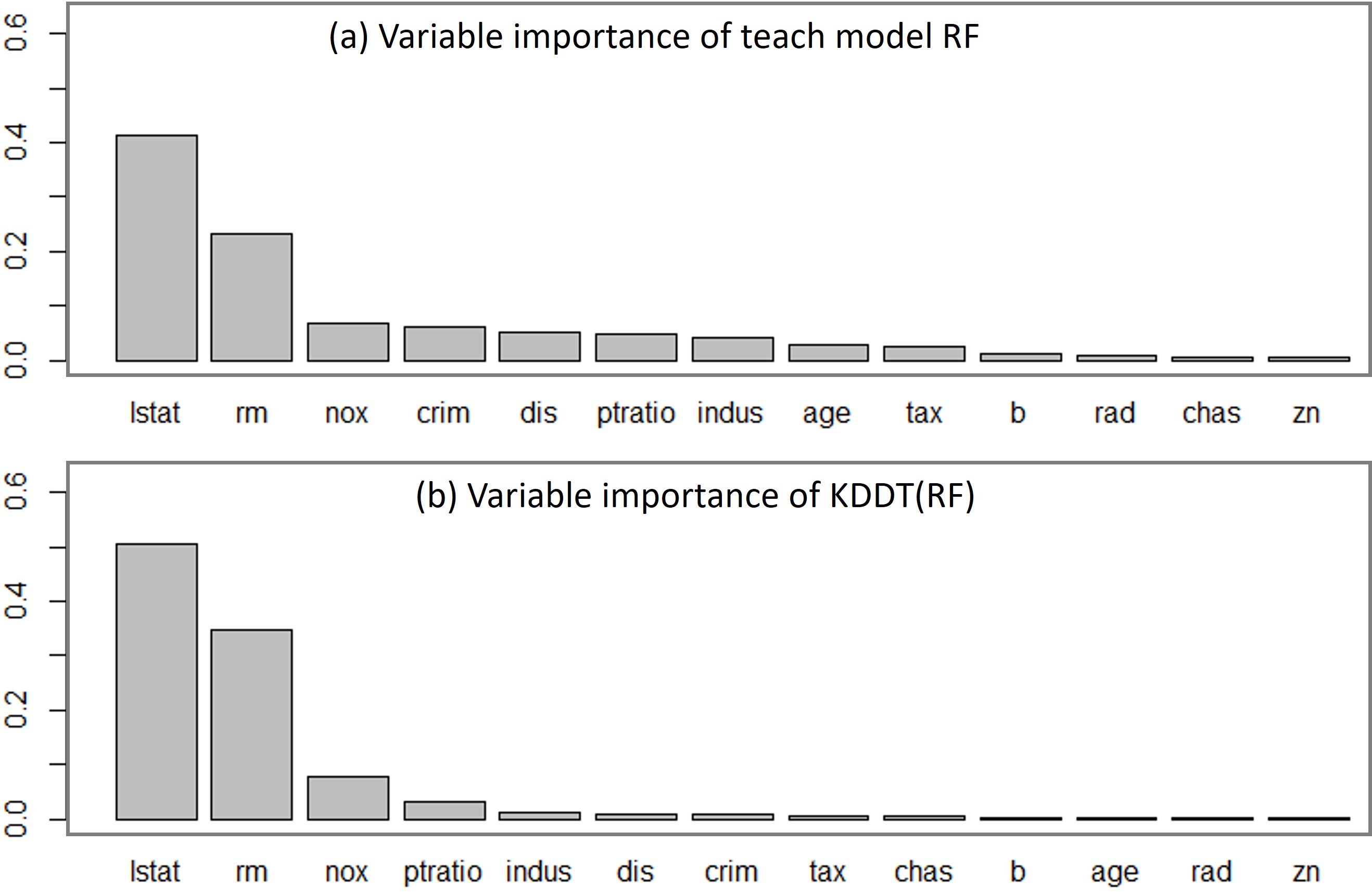}
\caption{\small Comparison of variable importance between the teacher Random Forest and the student KDDT(RF).}
\label{fig:BHvarImp}
\end{figure*}

\begin{figure*}[ht]
\centering
\includegraphics[width=.8\textwidth]{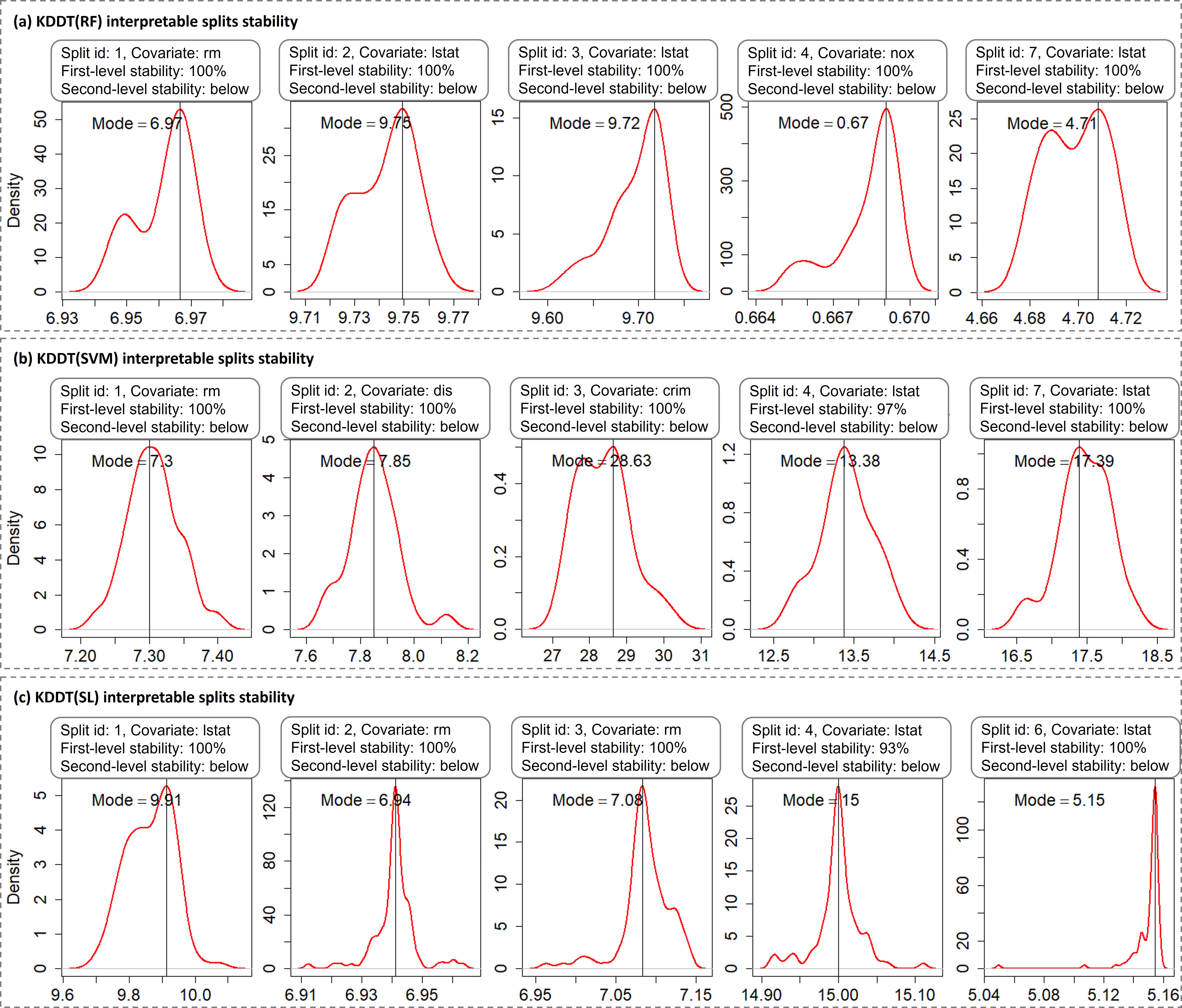}
\caption{\small Two-level stability of the interpretable splits in KDDT(RF), KDDT(SVM) and KDDT(SL).}
\label{fig:BHstability}
\end{figure*}

\begin{table*}[ht]
\renewcommand{\arraystretch}{1}
\caption{\label{tb:whasVars}Variables and short descriptions}
\small
\centering
\fbox{%
\begin{tabularx}{.7\textwidth}{|l|X|}
  \hline
  Variable & \multicolumn{1}{c|}{Short descriptions}  \\
  \hline \hline
  age & Age (per chart) (years). \\
  \hline
  sex & Sex. 0 = Male. 1 = Female. \\
  \hline
  cpk & Peak cardiac enzyme (iu). \\
  \hline
  sho & Cardiogenic shock complications. 1 = Yes. 0 = No. \\
  \hline
  chf & Left heart failure complications. 1 = Yes. 0 = No. \\
  \hline
  miord & MI Order. 1 = Recurrent. 0 = First. \\
  \hline
  mitype & MI Type. 1 = Q-wave. 2 = Not Q-wave. 3 = Indeterminate. \\
  \hline
  lenstay & Days in hospital. \\
  \hline
  lenfol & Total length of follow-up from hospital admission (days). \\
  \hline
  fstat & Status as of last follow-up. 1 = Dead. 0 = Alive. \\
  \hline
\end{tabularx}}
\end{table*}

\begin{figure*}[ht]
\centering
\includegraphics[width=.95\textwidth]{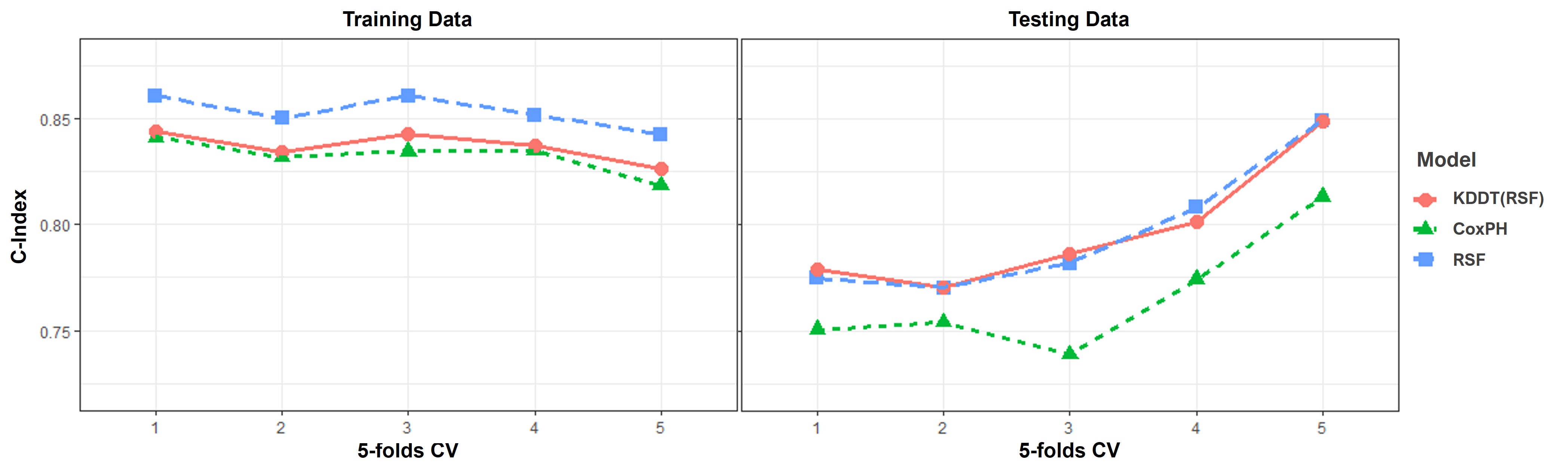}
\caption{\small Comparison of prediction accuracy among CoxPH, RSF, and KDDT(RSF) on the training dataset (left) and testing dataset (right). Note that RSF is the teacher model and KDDT(RF) is the student model. A higher C-index indicates superior performance in prediction. Notably, KDDT(RSF) surpasses its teacher model RSF on the first and third folds of the testing data. This is because KDDT(RSF), being an approximation of RSF, might relieve overfitting on the testing data.}
\label{fig:realWhasCompare}
\end{figure*}

\FloatBarrier

\end{appendix}

\newpage


\section{Acknowledgments}
The authors appreciate the helpful discussions with Dr. Wei-Ying Lou and the editorial assistance from Mrs. Jessica Swann.


\newpage

\bibliographystyle{chicago}

\bibliography{DDT}
\end{document}